\documentclass[a4paper,11pt]{scrarticle}

\usepackage[
    backend=biber,
    style=alphabetic,
    sorting=ynt,
    maxalphanames=4,
    minalphanames=3,
    maxbibnames=99,
    maxcitenames=99,
    url=false,
]{biblatex}
\AtEveryBibitem{\clearname{editor}}

\usepackage[margin=1in]{geometry}

\usepackage[T1]{fontenc}
\usepackage{lmodern}
\usepackage[english]{babel}
\usepackage{microtype}

\usepackage{amsmath,amssymb,amsthm}
\usepackage{csquotes}
\usepackage{hyperref}
\usepackage{caption}
\usepackage{mathtools}
\usepackage{xfrac}
\usepackage[dvipsnames]{xcolor}
\usepackage{graphicx}
\newcommand\linenumbers{}
\newcommand\nolinenumbers{}
\usepackage{orcidlink}

\usepackage[capitalise,noabbrev]{cleveref} %

\newcommand{\compclass}[1]{\ensuremath{\mathsf{#1}}}

\newcommand{\ccVP}{\compclass{VP}}
\newcommand{\ccVNP}{\compclass{VNP}}
\newcommand{\ccVPe}{\compclass{VF}}
\newcommand{\ccVF}{\compclass{VF}}
\newcommand{\ccVFk}[1]{\compclass{VF}_{#1}}
\newcommand{\ccVBP}{\compclass{VBP}}
\newcommand{\ccVBPk}[1]{\compclass{VBP}_{#1}}

\newcommand{\ccVFbar}{\compclass{\overline{VF}}}
\newcommand{\ccVBPkbar}[1]{\compclass{\overline{VBP}}_{#1}}

\newcommand{\ccMonVP}{\compclass{mVP}}
\newcommand{\ccMonVNP}{\compclass{mVNP}}
\newcommand{\ccMonVPe}{\compclass{mVF}}
\newcommand{\ccMonVF}{\compclass{mVF}}
\newcommand{\ccMonVFk}[1]{\compclass{mVF}_{#1}}
\newcommand{\ccMonVBP}{\compclass{mVBP}}
\newcommand{\ccMonVBPk}[1]{\compclass{mVBP}_{#1}}

\newcommand{\IN}{\mathbb{N}}
\newcommand{\IF}{\mathbb{F}}
\newcommand{\IR}{\mathbb{R}}

\newcommand{\bX}{\mathbf{X}}
\newcommand{\bY}{\mathbf{Y}}

\newcommand{\cC}{\mathcal{C}}
\newcommand{\eps}{\varepsilon}

\DeclareMathOperator{\supp}{supp}
\DeclareMathOperator{\poly}{poly}

\newcommand{\imm}[2]{\operatorname{imm}_{#1, #2}} %
\newcommand{\per}[1]{\operatorname{per}_{#1}}

\newcommand{\ffull}[1]{f^{(#1)}}
\newcommand{\fpartial}[3]{f^{(#3)}_{#1 \to #2}} %

\makeatletter

\newcommand{\Deltatimesaux}[2]{%
  \vcenter{\hbox{%
    \ooalign{%
      \hfil$#1\Delta$\hfil\cr\kern -0.5pt
      \hfil\raisebox{0.28ex}{\scalebox{0.4}{$#1\times$}}\hfil\cr
    }%
  }}%
}

\newcommand{\Deltaplusaux}[2]{%
  \vcenter{\hbox{%
    \ooalign{%
      \hfil$#1\Delta$\hfil\cr\kern -0.5pt
      \hfil\raisebox{0.28ex}{\scalebox{0.4}{$#1+$}}\hfil\cr
    }%
  }}%
}
\makeatother

\newcommand{\growth}{{\mathrm{Gr}}}

\newcommand{\dmult}{\ensuremath{d^{\kern -1pt \scalebox{0.8}{$\times$}}}}
\newcommand{\dadd}{\ensuremath{d^{\kern -1pt \scalebox{0.8}{$+$}}}}

\numberwithin{equation}{section}
\theoremstyle{plain}
\newtheorem{theorem}[equation]{Theorem}
\newtheorem{lemma}[equation]{Lemma}
\newtheorem{corollary}[equation]{Corollary}
\newtheorem{proposition}[equation]{Proposition}

\theoremstyle{definition}
\newtheorem{definition}[equation]{Definition}

\theoremstyle{remark}

\title{A Width Hierarchy of Monotone Algebraic Branching Programs}
\author{Julian Dörfler\thanks{Saarland Informatics Campus (SIC), Center for Quantum Technologies (QuTe), Saarland University, Germany, Email: \texttt{jdoerfler@cs.uni-saarland.de} ORCID: \orcidlink{0000-0002-0943-8282}\\ Funding: We gratefully acknowledge financial support from the "Transformationsprogramm Forschung und Wissenstransfer Saar" through the Center for Quantum Technologies (QuTe).} \and Alexander Rogovskyy\thanks{Saarland Informatics Campus (SIC), Saarbrücken Graduate School of Computer Science, Saarland University, Germany. Email: \texttt{rogovskyy@cs.uni-saarland.de} ORCID: \orcidlink{0009-0009-0941-7519}}}
\date{September 2026}

\begin{document}

\pagenumbering{gobble}
\maketitle

\begin{abstract}
    It is well known that $\ccVPe = \ccVBPk{3}$ (Ben-Or-Cleve'92), i.e.\ the width of any constant-width algebraic branching program can be reduced to width $3$ with only a polynomial overhead to size.
    We prove that this is not the case for the monotone equivalents.
    In particular, we prove that there is an infinite hierarchy 
    $\ccMonVBPk{2} \subsetneq \ccMonVBPk{3} \subsetneq \ldots$
    by showing that $\imm k n$ requires width $w$ mABPs to have depth at least $\left(\frac{k}{w-1}\right)^{\frac{1}{2}\left\lfloor n^{\sfrac{1}{(w-2)}} \right\rfloor - 1}$ whenever $3 \leq w < k$.
    We prove this lower bound using support-based techniques.
    
    Since this lower bound is exact instead of asymptotic, it allows some width separations even for superconstant width.
    Moreover, it allows us to show that the expressive power of poly-sized monotone formulas of depth $2d$ is \emph{not} reached by poly-sized mABPs of width $d-1$ but a width of $2d+1$ is sufficient.

    We contrast this to \emph{unbounded}-depth monotone formulas, which are proven to be stronger than constant-width mABPs, but weaker than $\log$-width mABPs and incomparable for most widths in between.
    
    Finally, we generalize Snir's framework (Snir'80) for showing lower bounds on the number of leaves of monotone formulas to monotone formulas of multiplication depth $\dmult$, giving a lower bound of $n \cdot k^{\dmult(n^{\sfrac{1}{\dmult}} - 1)}$ for the iterated matrix multiplication polynomial and $\left(\left(e+o(1)\right) \ln^{\circ(\dmult-1)}(n)\right)^n$ for the permanent.
\end{abstract}

\newpage
\pagenumbering{arabic}

\linenumbers

\section{Introduction}

Separations of different computational models have been a major objective in Algebraic Complexity Theory, the primary question being the relationship of the determinant and the permanent. In terms of complexity classes, the question resides in the relationship between the classes $\ccVBP$ and $\ccVNP$, one possible algebraic analog to the $\mathsf{P}$ vs. $\mathsf{NP}$ problem.

Since that question turned out to be far from understood, focus shifted in recent years towards specific restricted models. The focus of this paper are \emph{monotone} classes, that is, polynomials with non-negative coefficients. Historically, progress in the monotone world has been more fruitful, dating as far back as 1980 with \textcite{DBLP:conf/icalp/Snir80}, establishing the separation $\ccMonVF \subsetneq \ccMonVBP$ by showing that the iterated matrix multiplication polynomial $\imm k n$ requires monotone formulas with at least $(2k)^{\log(n)}$ leaves, refining techniques from a slightly earlier work by \textcite{DBLP:journals/mst/ShamirS80}.

In the following years, \textcite{DBLP:journals/jacm/JerrumS82} proved a superpolynomial lower bound on the size of a monotone algebraic circuit computing the permanent.

Much later, the groundbreaking results by \textcite{DBLP:conf/icalp/HrubesY16,DBLP:conf/stoc/Yehudayoff19} separated $\ccMonVBP$, $\ccMonVP$ and $\ccMonVNP$, resulting in the following known separations of classes:
\begin{equation*}
    \ccMonVF \subsetneq \ccMonVBP \subsetneq \ccMonVP \subsetneq \ccMonVNP.
\end{equation*}

In recent years, there has also been increasing attention to exploiting graph parameters (such as treewidth, pathwidth and related notions) to characterize which of these classes certain graph homomorphism polynomials belong to, see e.g.\ \cite{DBLP:journals/algorithmica/KomarathPR23,DBLP:conf/mfcs/BhargavCC025,DBLP:conf/mfcs/KomarathN26}.

Following this \enquote{rough} separation, one may also consider a single computational model under different parameters, for example the multiplication depth of a formula.
In this context, \textcite{DBLP:conf/focs/ChillaraEL018} proved superpolynomial separations between monotone circuits (and thus also monotone formulas) differing in depth, while \textcite{DBLP:conf/isaac/ArvindJS09} have proven a similar result for constant-\emph{width} monotone circuits. 
Moreover, \textcite{DBLP:conf/mfcs/BhargavCC025,DBLP:conf/mfcs/KomarathN26} have been able to give fine-grained size characterizations for graph homomorphism polynomials in monotone circuits (and monotone formulas respectively) of fixed depth.

A similar \enquote{internal hierarchy} has not been established for monotone algebraic branching programs (mABPs for short). The natural parameter here is the \emph{width} of an ABP. This choice of parameter is fairly intuitive: ABPs of a fixed width $w$ correspond to computations limited to $w$ registers.
We consider $\ccVBPk{w}$ to be the class of families of polynomials computable by polynomially sized width $w$ ABPs.
Similarly, $\ccMonVBPk{w}$ is the class of families of monotone polynomials computable by polynomially sized width $w$ mABPs.

In the non-monotone world, \enquote{adding a register} makes almost no difference, as long as the width is bounded: \textcite{DBLP:journals/siamcomp/Ben-OrC92} have proven $\ccVBPk{w} = \ccVF$ for all $w \geq 3$ while \textcite{DBLP:journals/cc/AllenderW16} have proven $\ccVBPk{2} \subsetneq \ccVBPk{3}$.
If approximations are allowed, \textcite{DBLP:journals/jacm/BringmannIZ18} have proven, even width $2$ is sufficient to reach the full computational power of $\ccVF$ (and even $\ccVFbar$), i.e.\ $\ccVFbar = \ccVBPkbar{w}$ for all $w \geq 2$ (over fields with characteristic different from $2$). 

\subsection{Our Contributions}

\begin{figure}
    \centering
    \hspace{-1cm}
    \resizebox{1\linewidth}{!}{%
         \usetikzlibrary{matrix, fit, positioning, calc, arrows.meta, decorations.pathmorphing}

\definecolor{ourResults}{HTML}{b01b2f}
 
 \newcommand{\elidedarrow}[4]{%
   \draw[#3] ($(#1.east)$) -- ($(#1.east)!0.3!(#2.west)$);
   \draw[dotted, thick, #3] ($(#1.east)!0.3!(#2.west)$) -- node[above]{#4} ($(#1.east)!0.7!(#2.west)$);
   \draw[->, #3] ($(#1.east)!0.7!(#2.west)$) -- (#2.west);%
 }

\begin{tikzpicture}[
    >=Stealth,
    box/.style   = {draw, rectangle, inner sep=4pt},
    squig/.style = {->, decorate,
                    decoration={snake, amplitude=0.4mm, segment length=2mm, pre length = 1.5mm, post length=1.5mm}},
  ]

  \matrix (m) [matrix of math nodes, column sep=1.5em, row sep=1.5cm, nodes={anchor=center}] {
    |[alias=mVBP2]| \ccMonVBPk{2}   &
    |[alias=mVBP3]| \ccMonVBPk{3}   &
    \ldots          &
    |[alias=mVBPk]|  \ccMonVBPk{k-1} &
    \ldots          &
    |[alias=mVBP2k]| \ccMonVBPk{2k+1}&
    |[alias=ConstantWidthEtc]| \ldots &[-1em]
    |[alias=mVBPO1]| =\ccMonVBPk{O(1)} \\
    |[alias=mVF2]| \ccMonVFk{2}    &
    |[alias=mVF3]| \ccMonVFk{3}    &
    |[alias=mVF4]| \ccMonVFk{4}    &
    \ldots          &
    |[alias=mVF2k]| \ccMonVFk{2k}   &
    |[alias=ConstantDepthEtc]| \ldots & &[-1em] 
    |[alias=mVFO1]| =\ccMonVFk{O(1)} \\
  };

  \node[box, fit=(mVBP2) (ConstantWidthEtc) (mVBP2 -| mVF2.west) (ConstantWidthEtc -| ConstantDepthEtc.east)] (boxA) {};
  \node[box, fit=(mVF2) (ConstantDepthEtc) (mVF2 -| mVBP2.west) (ConstantDepthEtc -| ConstantWidthEtc.east)] (boxB) {};

  \foreach \x [evaluate=\x as \xx using int(\x+1)] in {1,...,6} {
     \draw[->, ourResults] (m-1-\x) -- node[above]{\tiny 1} (m-1-\xx);
  }
  \foreach \x [evaluate=\x as \xx using int(\x+1)] in {1,...,5} {
     \draw[->] (m-2-\x) -- node[above]{\tiny 2} (m-2-\xx);
  }

  \draw[squig, ourResults] (mVF2k |- boxB.north) -- node[below, pos = 0.85]{\tiny 3} (mVBPk |- boxA.south);
  \draw[->, ourResults]    (mVF2k |- boxB.north) -- node [below, pos = 0.85]{\tiny 4} (mVBP2k |- boxA.south);
  \draw[squig] (mVBP2 |- boxA.south) -- node[below, pos = 0.15]{\tiny 5} (mVFO1 |- boxB.north);

  \draw[->, ourResults] (mVFO1 |- boxB.north) -- node[left]{\tiny 4\textcolor{black}{,5}} (mVBPO1);

  \matrix (m2) [matrix of math nodes, nodes in empty cells, column sep=1.3em, row sep=1.5cm, nodes={anchor=center}, below = 0.6 of m ] {
    |[alias=mVBPO1Second]| \ccMonVBPk{O(1)}   &[0.7cm]
    |[alias=mVBPg]| \ccMonVBPk{g(n)}   &
    |[alias=mVBP2g]|\ccMonVBPk{(1+\eps)g(n)}  &
    |[alias=SlowGrowingWidthEtc]| \ldots      &[0.7cm]
    |[alias=mVBPlogn]| \ccMonVBPk{O(\log n)} &
    |[alias=mVBP2ToTheRootLogN]| \ccMonVBPk{2^{\sqrt{\log n}}} &
    |[alias=mVBP]| \ccMonVBP \\
    |[alias=mVFO1Second]| \ccMonVFk{O(1)} & 
    & & & \\
  };
  \node (mVF) at ($(m2-2-2)!0.5!(m2-2-4)$) {$\ccMonVFk{O(\log n)} \overset{7}{=} \ccMonVF$};

  \node[box, fit=(mVBPg) (SlowGrowingWidthEtc) ] (boxC) {};

  \draw[->, ourResults] (mVFO1Second) -- node [left]{\tiny 4\textcolor{black}{,5}} (mVBPO1Second |- boxC.south);

  \draw[->, ourResults] (mVBPO1Second) -- node [above]{\tiny 6} (mVBPg -| boxC.west);

  \draw[->, ourResults] (SlowGrowingWidthEtc -| boxC.east) -- node [above]{\tiny 6} (mVBPlogn);

  \draw[->, ourResults] (mVBPg) -- node [above]{\tiny 6} (mVBP2g);

  \draw[->, ourResults] (mVBP2g) -- node [above]{\tiny 6} (SlowGrowingWidthEtc);

  \draw[->, ourResults] (mVBPlogn) -- node [above]{\tiny 11} (mVBP2ToTheRootLogN);

  \draw[->, ourResults] (mVBP2ToTheRootLogN) -- node [above]{\tiny 11} (mVBP);

  \draw[->] (mVFO1Second) -- node [above]{\tiny 7} (mVF);

  \draw[->, ourResults] (mVBPO1Second) -- node [above]{\tiny 8} (mVF);

  \draw[->, ourResults] (mVF) -- node [above]{\tiny 10} (mVBPlogn);

  \draw[squig, <->, ourResults] (mVF) -- node [right]{\tiny 9} (mVF |- boxC.south);

  \draw[->] (mVF.east) -- node [above]{\tiny 12} (mVBP);

\end{tikzpicture}%
    }
    \captionsetup{labelsep=newline}
    \caption[]{\parbox[t]{0.9\linewidth}{%
    Diagram of known and new relations between classes in the monotone world. A straight arrow represents a \emph{strict} inclusion. A squiggly arrow represents a non-inclusion. Novel results established in this paper are drawn in red. Here, $k$ is a natural number and $g$ is a placeholder for any function in $o\left(\frac{\log n}{(\log\log n)^2}\right)$. The strict inclusions represented by (\textsuperscript{6}) remain valid even for $g(n) \in o\left(\frac{\log n}{\log\log n} \right)$, but then the incomparability (\textsuperscript{9}) is not proven.

    \begin{minipage}[t]{0.28\linewidth}
        \textsuperscript{1} \cref{cor:main_hierarchy_const}\\
        \textsuperscript{2} \cite{DBLP:conf/stoc/KlawePPY84,DBLP:conf/focs/ChillaraEL018} \\
        \textsuperscript{3} \cref{thm:ABPkNotInFormula2k}\\
        \textsuperscript{4} \cref{cor:formula_to_almost_same_width_abp}
    \end{minipage}\hfill
    \begin{minipage}[t]{0.33\linewidth}
        \textsuperscript{5} \cite{DBLP:journals/siamcomp/ChillaraL019}, \cref{cor:imm2_not_in_constant_depth_formulas}\\
        \textsuperscript{6} \cref{cor:main_hierarchy_fns}\\
        \textsuperscript{7} \cite{DBLP:journals/siamcomp/ChillaraL019,DBLP:conf/stoc/TavenasL022} \\
        \textsuperscript{8} \cref{thm:poly_in_mVF_but_not_in_slow_growing_width_mVBP}
    \end{minipage}\hfill
    \begin{minipage}[t]{0.33\linewidth}
        \textsuperscript{9} \cref{cor:mVBP_k_not_in_mVF_if_unbounded}, \cref{thm:poly_in_mVF_but_not_in_slow_growing_width_mVBP} \\
        \textsuperscript{10} \cref{prop:mVF_in_mVBP_O_logn} \\
        \textsuperscript{11} \cref{cor:coarse_abp_separation_logf_vs_logg} \\
        \textsuperscript{12} \cite{DBLP:conf/icalp/Snir80}
    \end{minipage}
    }}
    \label{fig:hierarchy}
\end{figure}

The completeness results of $\ccVBPk{3}$ (and $\ccVBPkbar{2}$) crucially depend on cancellations.
The natural question is then: What does \enquote{adding a register} change in the monotone world? Do we have a completeness theorem starting from a certain width in a similar fashion? If yes, what is that width and if no, how do these \enquote{lower} complexity classes and hierarchies (below $\ccMonVP$) relate to each other?

In this paper, we complete the picture in the monotone world in several ways.
A graphical representation of our results can be found in \cref{fig:hierarchy}.

\minisec{A Hierarchy of Fixed-width mABPs}{%
We establish a hierarchy of fixed-width mABPs. More precisely, we show $\ccMonVBPk{k} \subsetneq \ccMonVBPk{k+1}$ for all $k \geq 2$, in addition to the obvious $\ccMonVBPk{1} \subsetneq \ccMonVBPk{2}$. This establishes a hierarchy similar to the known hierarchy for bounded-depth formulas. The main technical work is done in \cref{sec:main_separation}, resulting in an explicit (non-asymptotic) lower bound in \cref{thm:main_separation}.
}

\minisec{Separations for Superconstant-width mABPs}{%
Slightly stronger yet, we obtain more fine-grained conditions for functions $f$ and $g$, under which $\ccMonVBPk{f(n)}$ and $\ccMonVBPk{g(n)}$ are superpolynomially separated. The precise conditions are stated in \cref{cor:main_hierarchy_fns}; a possible regime is for instance $f \in o(\frac{\log n}{\log \log n})$ (in fact, $f \in \Theta(\frac{\log n}{\log\log n})$ is also permitted as long as the hidden constant is $< 1$) and $g(n)$ satisfying $g(n) \geq c f(n)$ for $c >1$.
}

\minisec{Separations in the Coarse Regime}{%
Using a different technique, we obtain $\ccMonVBPk{f(n)} \subsetneq \ccMonVBPk{g(n)}$ for $\log f(n) \in o(\log g(n))$ (\cref{cor:coarse_abp_separation_logf_vs_logg}).
This is a slightly unusual condition and satisfied for instance with $f(n) = \log(n)$ and $g(n) = 2^{\sqrt{\log n}}$ or $f(n) =  2^{\sqrt{\log n}}$ and $g(n) = \poly(n)$.
}

\minisec{Refining the Relation of $\ccMonVF$ to the Monotone Hierarchy}{%
While $\ccMonVF \subsetneq \ccMonVBP$ has been long since known, in \cref{sec:placement_of_mVF} we are able to refine the relation to
\begin{equation*}
    \ccMonVBPk{O(1)} \subsetneq \ccMonVF \subsetneq \ccMonVBPk{O(\log n)}
\end{equation*}
while for $k \in o\left(\frac{\log n}{(\log \log n)^2}\right)$, $\ccMonVF$ and $\ccMonVBPk{k(n)}$ are incomparable.
}

\minisec{Connecting Bounded-width mABPs and Bounded-depth Formulas}{%
We obtain connections between the fixed-width mABP hierarchy and the fixed-depth monotone formula hierarchy. 
\begin{itemize}
    \item There is a polynomial family computable by poly-sized monotone depth-$2k$ formulas but not computable by poly-sized mABPs of width $k-1$. (\cref{thm:ABPkNotInFormula2k})
    \item However, any polynomial family computable by poly-sized monotone depth-$k$ formulas \emph{can} be computed by poly-sized mABPs of width $2\lceil\frac{k}{2} \rceil + 1$.(\cref{cor:formula_to_almost_same_width_abp})
    \item There is a polynomial family computable by a width $2$ poly-sized mABP but not computable by any monotone formula of bounded depth and polynomial size (\cref{cor:imm2_not_in_constant_depth_formulas}). This recovers a known result by \textcite{DBLP:journals/siamcomp/ChillaraL019} using different techniques.
\end{itemize}
From the perspective of monotone depth-$k$ formulas, this means that even though a width of roughly $\frac{1}{2}k$ is not enough, at a width of around roughly $k$, mABPs can capture the expressiveness.

On the other hand, for any fixed width $w$, there are infinitely many depths $d$ such that $\ccMonVBPk{w}$ and $\ccMonVFk{d}$ are incomparable, where $\ccMonVFk{d}$ is the class of families of polynomials computable by monotone formulas of depth $d$ and polynomial size.
}

\minisec{A Generalization of Snir's Framework for \emph{Bounded-Depth} Formulas.}{%
In \cite{DBLP:conf/icalp/Snir80}, Snir has developed lower-bound techniques for monotone formulas. In \cref{sec:snir}, we generalize this technique to work for monotone formulas of bounded depth. 

This allows us to show that a monotone multiplication-depth $\dmult$ formula computing the iterated $k \times k$ matrix multiplication polynomial of $n$ matrices needs at least $n \cdot k^{\dmult(n^{\sfrac{1}{\dmult}} - 1)}$ leaves (\cref{thm:depth_snir_imm}).
Moreover, any monotone depth $d$ and multiplication-depth $\dmult$ circuit for it requires size at least $n \cdot k^{\frac{\dmult}{d}(n^{\sfrac{1}{\dmult}} - 1)}$ (\cref{cor:circuit_depth_snir_imm}).
This generalizes a result by \cite{DBLP:journals/siamcomp/ChillaraL019} to $k > 2$ and improves upon an earlier $k^{\Omega\left(\frac{1}{d} n^{\sfrac{1}{\lceil \sfrac{d}{2} \rceil}}\right)}$ bound for monotone circuits by \cite{JAJA19851155}. 

As a further application, we show that the optimal size of a monotone multiplication-depth $\dmult$ formula computing the $n \times n$ permanent is $\left(\left(e+o(1)\right) \ln^{\circ(\dmult-1)}(n)\right)^n$ (\cref{thm:depth_snir_per}).

}

\minisec{A Width-Depth Tradeoff for mABPs}{%
The width-hierarchy shows that mABPs do not in general admit a width reduction while keeping polynomial size. However, it might occasionally be useful to allow a superpolynomial blow-up in size; the question then becomes how large that blow-up needs to be, given that we do have an explicit lower bound in \cref{thm:main_separation}.

In \cref{thm:narrow-ABPs-width-to-depth}, we show such a tradeoff:  the iterated $k \times k$ matrix multiplication of $n$ matrices can be computed by an mABP of width $w$ and depth at most $n \cdot \left\lceil \sfrac{k}{\left\lfloor\frac{w-2}{(2r-1)}\right\rfloor}\right\rceil^{r(\lceil n^{\sfrac{1}{r}} \rceil - 1)}$, where $r$ can be chosen between $1$ and $\lfloor \frac{w-1}{2} \rfloor$.
}

\section{Preliminaries}
\label{sec:prelims}

\subsection{Basic Notions}

Let $\IF$ be a field. For a polynomial $f \in \IF[x_1, \dots, x_m]$, the \emph{support} $\supp(f)$ is the set of its monomials with non-zero coefficients. By convention, $\supp(0) = \{\}$. For $\IF = \IR$, we call $f$ \emph{monotone} if every monomial in $\supp(f)$ has positive coefficient.
A polynomial $f \in \IF[x_1, \dots, x_m]$ is called \emph{homogeneous} if all monomials in its support have the same degree.

Let $\bY$ be a set of variables. A polynomial $f \in \IF[\bY]$ is \emph{set-multilinear} with respect to a partition $\bY = \bigsqcup_i \bY_i$ if every monomial in $\supp(f)$ has at most one variable from each block $\bY_i$.

A \emph{$p$-family} $(f_n)_{n\in \IN}$ (often also just denoted $(f_n)$) is a sequence of polynomials such that the number of variables and the degree of $f_n$ are both polynomially bounded in $n$. Generally speaking, algebraic complexity classes are defined as sets of $p$-families.

We say that a polynomial $f$ \emph{is a projection of} $g$ if we can write $f = g(a_1, \dots, a_m)$, where each $a_i$ is either a field element or a variable.
In the monotone world, the analog are \emph{monotone projections}, which are defined similarly except that each $a_i$ is either an element of $\IR_{\geq 0}$ or a variable. We will henceforth only say \enquote{projections} without specifying whether we mean monotone or not if it is clear from the context.

The logarithm $\log(x)$ by default is with respect to base $2$, while $\ln(x)$ is the natural logarithm.
We define the $i$-fold function application of some $f: \IR \to \IR$ as $f^{\circ i}$, i.e.\ $f^{\circ 0}(x) := x$ and $f^{\circ i}(x) = f^{\circ (i-1)}(f(x))$.

\subsection{Algebraic Branching Programs}

An \emph{algebraic branching program} (ABP) $A$ is a layered directed graph where edges only exist between consecutive layers.
The edges are labeled by affine linear forms over some field $\IF$, in our case, $\IF = \IR$.
Its first layer contains a single node $s$, called the \emph{source} and its last layer contains a single node $t$, called the \emph{sink}. We will refer to $s$ and $t$ when the ABP in question is clear from the context.

An ABP $A$ computes the polynomial $\ffull{A}$ given by summing over the polynomials computed by all $s$-$t$ paths.
The polynomial computed by a single $s$-$t$ path is the product of the labels on edges along the path.
The \emph{width} of $A$ is the number of nodes in its largest layer and the \emph{depth} is the number of edges in every $s$-$t$ path, i.e.\ the number of layers minus one. The \emph{size} of $A$ is its number of nodes.
We denote by $\fpartial u v A$ the sum of the polynomials of all $u$-$v$ paths in $A$, or, equivalently, the polynomial computed by the sub-ABP in $A$ with source $u$ and sink $v$.

If all edge labels in an ABP only contain coefficients from $\IR_{\geq 0}$, it is called a \emph{monotone} algebraic branching program (mABP).
The polynomial computed by an mABP is always monotone (has coefficients in $\IR_{\geq 0}$).

Our main focus is on the iterated matrix multiplication polynomial
\begin{equation*}
    \imm k n := (\bX^{(1)} \bX^{(2)} \cdots \bX^{(n)})_{1,1},
\end{equation*}
where the $\bX^{(i)} = (x^{(i)}_{\kappa, \lambda})_{\kappa,\lambda}$ are $k \times k$ matrices of variables.
It is naturally computed by a \enquote{fully connected} mABP of depth $n$ and width $k$, where an edge from vertex $\kappa$ in layer $i$ to a vertex $\lambda$ in layer $i+1$ is assigned the label $x^{(i)}_{\kappa, \lambda}$.
The iterated matrix multiplication is very structured: it is homogeneous, monotone and set-multilinear with respect to the variables in the matrices $\bX^{(1)}, \ldots, \bX^{(n)}$, i.e.\ every monomial has the form $x^{(1)}_{\kappa_1, \lambda_1} x^{(2)}_{\kappa_2, \lambda_2} \cdots x^{(n)}_{\kappa_n, \lambda_n}$.

Conversely, a monomial of that form is part of the support of $\imm k n$ iff $\kappa_1 = \lambda_n = 1$ and $\kappa_{i+1} = \lambda_{i}$ for all $i \in \{1, \ldots, n-1\}$.

We consider $\ccVBP$ to be the class of all $p$-families $(f_n)$ computable by ABPs of polynomial size in $n$ while $\ccVBPk{w}$ is the class of all $p$-families $(f_n)$ computable by width $w$ ABPs of polynomial size in $n$.
If $w$ is a function in $n$, then $\ccVBPk{w(n)}$ denotes the class of all $p$-families $(f_n)$, s.t.\ $f_n$ is computable by a width $w(n)$ ABP of polynomial size in $n$ for all $n \in \IN$.
If $W$ is some set of functions in $n$, then $\ccVBPk{W} = \bigcup_{w \in W}\ccVBPk{w(n)}$.
For example, we have $\ccVBPk{\poly(n)} = \ccVBP$.

Similarly, we define $\ccMonVBP$, $\ccMonVBPk{w}$, $\ccMonVBPk{w(n)}$ and $\ccMonVBPk{W}$ using mABPs instead of ABPs.

\subsection{Algebraic Formulas}

An algebraic formula $A$ (or formula for short) over a field $\IF$ is a rooted tree, where leaves are labeled by scalar multiples of variables\footnote{Variations exist where some authors instead of allowing multiples of variables allow scalar edge weights. This makes no difference and leads to equivalent models. We will allow multiples of variables although in practice, our coefficients will be $1$.} or field elements and inner nodes are either addition gates or multiplication gates. The \emph{fan-in} of a gate is the number of children. Every gate computes a polynomial over $\IF$ in the natural way; the polynomial computed by $A$ is the polynomial computed by its root. We call a formula \emph{monotone}, if all occurring constants are non-negative. The number of gates is the \emph{size} of the formula\footnote{Alternative, but polynomially related definitions (such as the number of edges) are possible.}.

The \emph{depth} $d$ of a formula is the maximal number of edges between the root and a leaf.
We will frequently consider families of formulas with bounded depth. In this regime, the fan-in is usually left unbounded to allow the formula to produce arbitrary-degree polynomials. Consequently, we may collapse consecutive addition (and similarly consecutive multiplication) gates in a path and assume that, on any path from the root, the addition and multiplication gates alternate. From now on we will w.l.o.g.\ assume that all our bounded-depth formulas have this alternating form.

We define the notion of \emph{addition depth} $\dadd$ as the maximum number of addition gates on the path between the root and a leaf. The \emph{multiplication depth} $\dmult$ is defined equivalently. Note that $\dadd$ and $\dmult$ differ by at most $1$ and that $\dadd + \dmult = d$.

For instance, the formula corresponding to $(x_1 x_2 + x_3 x_4) x_5$ has $d = 3$, $\dadd = 1$ and $\dmult = 2$.

The class $\ccVF$ consists of all $p$-families computable by formulas with polynomial size. Analogous to bounded-width ABPs, we allow functions $d(n)$ in the subscript: $\ccVFk{d(n)}$ is the class of $p$-families computable by formulas of bounded depth $d(n)$ and polynomial size. For a set $D$ of functions, $\ccVFk{D}$ is similarly defined as $\bigcup_{d\in D} \ccVFk{d(n)}$. The monotone classes $\ccMonVF$ and the corresponding versions with a subscript are defined by restricting to monotone formulas.

\section{Proof Overview}
\label{sec:proof_overview}

To separate $\ccMonVBPk{w}$ from $\ccMonVBPk{w+1}$, we show that it takes a width $w$ mABP a superpolynomial depth $d$ to compute the family of polynomials $f_n = \imm {w+1} n$.
To gain an understanding into our lower bound, we start by looking at \cref{fig:basic_decompose} to see what an actual width $w$ mABP $A$ for $f_n$ looks like:

\begin{figure}
    \centering
    \includegraphics[width=\linewidth]{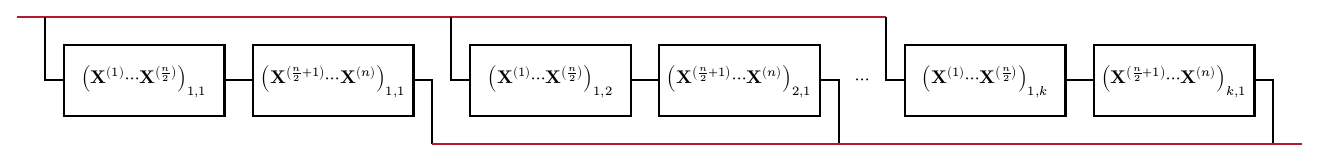}
    \caption{%
        The basic recursive structure for computing $\imm k n$ using a small width mABP.
        Once width $3$ is reached, the smaller iterated matrix multiplications just degenerate to individual monomials.
        All edges on the long paths at the top and bottom are labeled $1$.
        For space reasons, only decompositions into two parts are shown.
        To reach the required width $w$, one would usually have $\approx n^{\sfrac{2}{(w-1)}}$ parts, leading to a total of $\approx n^{\sfrac{2}{(w-1)}} \cdot w ^ {n^{\sfrac{2}{(w-1)}}-1}$ recursive iterated matrix multiplication blocks.
        See \cref{lem:imm_to_depth_formula,lem:abpsum,lem:abpprod} for the details.
    }
    \label{fig:basic_decompose}
\end{figure}

Essentially we have two long constant paths (that is, paths labeled by field elements, in this case $1$), one starting at the source of $A$ and one ending at the sink of $A$, with computations of width $w-2$ starting at the source-path, and ending at the sink-path.
Each of the smaller computations computes some iterated matrix multiplication itself.

It turns out, this general structure is not a coincidence:
Any width $w$ mABP computing $f_n$ is forced to have a source-path and a sink-path and all true computations happen between them, at a smaller width, allowing us to decompose $f_n$ as a sum of polynomials $g_1, \ldots, g_{\poly(d)}$, computed by smaller width mABPs of depth at most $d$ (\cref{lem:widthreduce}).
Assume for now that we could somehow guarantee that, after a projection, one of $g_1, \ldots, g_{\poly(d)}$ computes an iterated matrix multiplication $\imm {w+1} m$ with $m = \poly(n)$.
Then we could recursively apply the decomposition until we reach the point where a width $2$ mABP has to compute some $\imm {w+1} m$ with $m = \poly(n)$.
Here, however, we reach an impossibility result: width $2$ mABPs cannot compute $\imm k m$ at all for $k \geq 3$.
Even more, they can at most compute $2^m$ of its monomials 
(\cref{cor:2vs3}).
Using this quantitative result we can then derive a lower bound on $d$ (\cref{thm:main_separation}).

All of this hinges on the hope that, for one of the $g_i$, we can guarantee the existence of a projection to some $\imm {w+1} m$.
Unfortunately, we do not believe this claim to be true.
Fortunately, however, the recursion doesn't actually require a full $\imm {w+1} m$, but only some polynomial $h$ with $\supp(h) \subseteq \supp(\imm {w+1} m)$ and at least some fraction of the monomials.
After a recursion step the resulting fraction of monomials might drop too low, which we then boost back up by projecting to a slightly smaller iterated matrix multiplication instead (\cref{lem:make_dense}).

\section{The Separation}
\label{sec:main_separation}
The goal of this section is to establish a hierarchy $\ccMonVBPk{2} \subsetneq \ccMonVBPk{3} \subsetneq \dots$.
We proceed in four steps.
The first step is a standard simplification, while the second step concerns the base case of what will become an inductive proof.

\subsection{Simplification}
In our subsequent proofs, we will often proceed by counting monomials. 
Ideally, every $s$-$t$-path would contribute at most one monomial to the resulting polynomial. 
Of course, this is not true, even in the monotone setting. 
We therefore introduce a restricted class of mABPs that will allow for more straightforward counting.
We call an mABP $A$ \emph{simple}, if all vertices are relevant to the computation, no two vertices in the same layer are linearly dependent and all edges are labeled either by constants or a multiple of a single variable.
In particular, we want that there is no vertex $v$ with $\fpartial{s}{v}{A} = 0$ or $\fpartial{v}{t}{A} = 0$ and, in every layer, there are no two vertices $u, v$ with $\fpartial{s}{u}{A} = c \cdot \fpartial{s}{v}{A}$ or $\fpartial{u}{t}{A} = c \fpartial{v}{t}{A}$ for some constant $c \neq 0$.
For a simple mABP $A$, it is easy to see that we can indeed bound $|\supp(\ffull{A})|$ by the number of $s$-$t$-paths in $A$. Fortunately, we can show that simplicity is not a real restriction.

\begin{lemma}
    \label{lem:simplicity}
    Let $A$ be a width $w$, depth $d$ mABP computing a non-zero polynomial $\ffull{A}$ with $\supp(\ffull{A}) \subseteq \supp(\imm{k}{n})$ for some $k, n$. Then there is a \emph{simple} mABP $A'$ with width $w$ and depth at most $d$ with $\ffull{A} = \ffull{A'}$.
\end{lemma}
The proof of \cref{lem:simplicity} is quite standard; we refer the interested reader to \cref{sec:simplify} for a full exposition.

\subsection{Separating Width 2 from 3}

Our initial starting point will be separating width $2$ mABPs from width $3$. It should be noted that separating width $1$ from $2$ is trivial (as the former produces $1$ monomial if it is simple, which we can assume, whereas $\imm 2 n$ has $2^{n-1}$ monomials).
This will serve as the base case for our main separation theorem.

We start with a small technical tool:
\begin{lemma}
    \label{lem:switching}
    Let $f_1, f_2, g_1, g_2$ be monomials with $\{f_1 g_1, f_1 g_2, f_2g_2\} \subseteq \supp(\imm k n)$.
    Then, $f_2g_1 \in \supp(\imm k n)$.
\end{lemma}
\begin{proof}
    Assume for the sake of contradiction that $f_2g_1 \notin \supp(\imm k n)$.
    Due to set-multilinearity of $f_1 g_1$, $f_1 g_2$ and $f_2g_2$ we know that $f_2g_1$ is also set-multilinear.
    The only way we can have $f_2g_1 \notin \supp(\imm k n)$ is thus by having $f_2g_1 = \prod_{i=1}^{n}x^{(i)}_{\mu_i, \nu_i}$ with $\nu_{i} \neq \mu_{i+1}$ for some $i$.
    Moreover, $x^{(i)}_{\mu_i, \nu_i}$ and $x^{(i+1)}_{\mu_{i+1}, \nu_{i+1}}$ cannot both be part of $f_2$ or $g_1$, otherwise either $f_2g_2$ or $f_1g_1$ would not be a monomial in $\imm k n$.
    W.l.o.g, let $x^{(i)}_{\mu_i, \nu_i}$ be part of $f_2$ and let $x^{(i+1)}_{\mu_{i+1}, \nu_{i+1}}$ be part of $g_1$.
    Then $f_1$ has to contain some variable $x^{(i)}_{\kappa, \mu_{i+1}}$ due to $f_1 g_1 \in \supp(\imm k n)$ and $g_2$ has to contain some variable $x^{(i+1)}_{\nu_{i}, \lambda}$ due to $f_2 g_2 \in \supp(\imm k n)$.
    However, then $f_1g_2$ contains both $x^{(i)}_{\kappa, \mu_{i+1}}$ and $x^{(i+1)}_{\nu_{i}, \lambda}$, while we have $\nu_i \neq \mu_{i+1}$, a contradiction to $f_1g_2 \in \supp(\imm k n)$.
\end{proof}

\begin{lemma}
    \label{lem:depthreduce}
    Let $f$ be a monotone polynomial computed by a width $2$ mABP with $\supp(f) \subseteq \supp(\imm k n)$ for some $k, n$.
    Then there is monotone polynomial $f'$ with $\supp(f) \subseteq \supp(f') \subseteq \supp(\imm k n)$ computed by some width $2$ mABP of depth $n$.
\end{lemma}
\begin{proof}
    Let $f$ be computed by a simple width $2$ mABP $A$ using \cref{lem:simplicity}.
    We split $A$ into connected components $\cC = \{C_1, \ldots, C_r\}$ only having inner edges labeled by constants.
    In each component $C_i$ all nodes $v$ compute polynomials $\fpartial s v A$ of the same degree $d_i$.
    
    Moreover, we claim that there are at most two components with the same degree.
    To see this, assume that $a, b, c$ are nodes on layers $\ell_a, \ell_b, \ell_c$ from distinct components computing the same degree polynomials, so in particular they are pairwise not connected (otherwise, there would be a degree jump along their connecting path).
    W.l.o.g, $\ell_a \leq \ell_b \leq \ell_c$. 
    Let $b'$ be the other node on the same layer as $b$.
    Since $a$ is not connected to $b$, there must be a (possibly empty) path from $a$ to $b'$ due to simplicity.
    Likewise, there is a path from $b'$ to $c$.
    But now $a$ and $c$ are connected, a contradiction.
    
    Now construct the mABP $A'$ as follows:
    Add one vertex $i$ for each component $C_i$ in layer $d_i$.
    Connect vertices $i$ and $j$ by an edge, weighted with the sum of the weights of all edges crossing from component $C_i$ to component $C_j$.
    Clearly, $A'$ is monotone, has width at most $2$ and has depth $n$.
    Moreover, any monomial generated by $A$, will also be generated by $A'$.
    It remains to show that any additionally generated monomials by $A'$ are in the support of $\imm k n$.
    Any such monomial corresponds to a path from $s$ to $t$ in $A$ that is allowed to move forwards arbitrarily and additionally backwards through constant edges.
    Iteratively, we see that each backwards section can only introduce desired monomials:
    Let $a$ and $b$ be the vertices at the start and end of a backwards section respectively.
    By simplicity, all of the polynomials $\fpartial s a A \cdot \fpartial a t A$, $\fpartial s b A \cdot \fpartial b t A$, and $\fpartial s b A \cdot \fpartial b a A \cdot \fpartial a t A$ are non-zero and all monomials contained within are in the support of $\imm k n$.
    By \cref{lem:switching} and using the fact that $\fpartial b a A$ is a non-zero constant, the polynomial $\fpartial s a A \cdot \fpartial b a A \cdot \fpartial b t A$ -- containing all monomials using this specific backwards section -- also only contains monomials in the support of $\imm k n$.
    Repeating this argument allows us to add one backwards section at a time, until we have finally formed the closure, represented by $A'$.
\end{proof}

\begin{corollary}
    \label{cor:2vs3}
    Width $2$ mABPs cannot compute $\imm 3 n$ for $n \geq 2$, a polynomial with $3^{n-1}$ monomials, at all.
\end{corollary}
\begin{proof}
    If a width $2$ mABP $A$ computes $\imm 3 n$ (a polynomial with $3^{n-1}$ monomials), \cref{lem:depthreduce,lem:simplicity} shows that there is a simple mABP $A'$ of width $2$ \emph{and depth $n$} computing it.
    However, $A'$ can produce at most $2^{n-1}$ monomials due to simplicity.
\end{proof}

Thus, we directly obtain a separation between width $2$ and width $3$ mABPs.
For the higher width separations we cannot hope for such a strong separation: width $3$ mABPs can already compute all monotone polynomials in a monomial-by-monomial fashion, albeit not with polynomial size (also see \cref{lem:abpsum,cor:formulatoabp}).

\subsection{Width-Reduction}

We will now aim to \enquote{decompose} a width $w$ mABP $A$ computing some larger $\imm k n$ (or a sufficiently high fraction of its support) into polynomially many mABPs of width $w-1$.
This will serve as the primary width-reduction gadget in our separation theorem.

\begin{figure}
    \centering
    \includegraphics[width=0.8\linewidth]{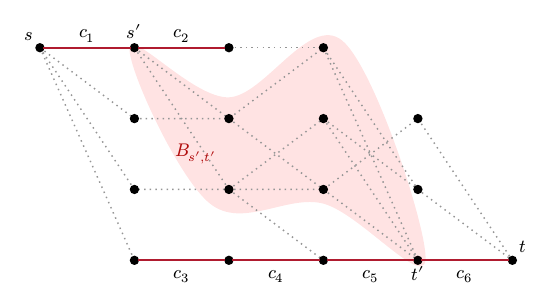}
    \caption{Schematic of a width $4$ mABP $A$ with \enquote{constant-paths} $V^s$ and $V^t$ in red. Their interior edge labels are constants $c_i$. The sub-mABP $B_{s', t'}$ is highlighted. The dotted paths illustrate possible edges in $A$. Note that $B_{s', t'}$ has width $3$. }
    \label{fig:constant-paths}
\end{figure}

The main idea is the observation that any such mABP must contain two distinguished paths $V^s$, $V^t$ where the edges are labeled by constants, along which the mABP can be decomposed (see also \cref{fig:constant-paths}).

\begin{lemma}[Decomposition Lemma]
    \label{lem:widthreduce}
    Let $3 \leq w < k$ and let $f$ be a monotone polynomial with $\supp(f) \subseteq \supp(\imm k n)$ for some $n \in \IN$ and $|\supp(f)| > \frac{w}{k} \cdot |\supp(\imm k n)| = w \cdot k^{n-2}$.

    If $f$ is computed by a width $w$ mABP $A$ of depth $d$, then there are $r < d^2$
    mABPs $B_1, \dots, B_r$, each of width $w-1$ and depth $\leq d$, such that $f$ is a linear combination of $\ffull{B_i}$.
    
\end{lemma}
\begin{proof}
    Using \cref{lem:simplicity}, w.l.o.g., we assume $A$ to be simple.

    As a first step, we prove that every layer of $A$ contains a node $v$ with either $\fpartial s v A$ or $\fpartial v t A$ being a constant.
    Consider any node $v$ where both $\deg(\fpartial s v A)$ and $\deg(\fpartial v t A)$ are $> 0$.
    Due to $\supp(f) \subseteq \supp(\imm k n)$ we know that $f$ is set-multilinear with respect to the matrices $\bX^{(1)}, \ldots, \bX^{(n)}$.
    Combined with the monotonicity of $A$, this implies that $\fpartial s v A$ and $\fpartial v t A$ partition these matrices. Additionally, since we assume they are both non-constant, they are both assigned at least one matrix.
    Then, there must be some $\ell$, s.t.\ $\bX^{(\ell)}$ is assigned to $\fpartial s v A$ and $\bX^{(\ell+1)}$ is assigned to $\fpartial v t A$ or vice versa (in which case we can mirror $A$ to swap $s$ and $t$ to obtain the first variant).
    Let $\fpartial s v A$ contain the variable $x_{\iota,\kappa}^{(\ell)}$.
    Then $\fpartial v t A$ can only contain the variables $x_{\kappa,\lambda}^{(\ell+1)}$ for $\lambda \in \{1, \ldots, k\}$, otherwise we would obtain a monomial not present in $\supp(\imm k n)$.
    This lets us bound the number of monomials of $\fpartial s v A \cdot \fpartial v t A$ by $k^{n-2}$.
    Each layer of $A$ has at most $w$ nodes and every monomial computed by $A$ has to use some node of this layer.
    Thus, $A$ would be able to only compute a polynomial with $w \cdot k^{n-2}$ monomials, a contradiction to $|\supp(f)| > w \cdot k^{n-2}$.
    Therefore, every layer contains some node $v$ with either $\fpartial s v A$ or $\fpartial v t A$ being a constant.

    Now consider the \enquote{constant-paths} $V^s := \{v \mid \fpartial s v A \in \IR\}$ and $V^t := \{v \mid \fpartial v t A \in \IR\}$.
    Note that $V^s$ and $V^t$ are indeed (disjoint) paths, with $V^s$ starting in $s$ and $V^t$ ending in $t$. To see this, first observe that $V^s$ is closed under predecessors.
    In addition, any vertex in $V^s$ except $s$ has in-degree $1$ due to simplicity.
    Similar statements hold for $V^t$.
    Disjointness of $V^s$ and $V^t$ follows from monotonicity since $\imm k n$ does not contain a constant monomial.
    
    For $s' \in V^s$ and $t' \in V^t$ consider the mABP $B_{s',t'}$ obtained from $A$ by using $s'$ as a source, $t'$ as a sink and containing no other nodes from $V^s$ or $V^t$ (see \cref{fig:constant-paths}).
    Note that any $s$-$t$-path in $A$, after leaving $V^s$, cannot reenter it again since $V^s$ is closed under predecessors (and similarly for $V^t$).
    Thus, we get a decomposition 
    \begin{align*}
        f &= \sum_{\substack{s' \in V^s\\t' \in V^t}} \fpartial s {s'} A \cdot \fpartial {s'} {t'} {B_{s', t'}} \cdot \fpartial {t'} t A\\
        &= \sum_{\substack{s' \in V^s\\t' \in V^t}} c_{s',t'} \cdot \fpartial {s'} {t'} {B_{s', t'}}
    \end{align*}
    with $c_{s',t'} \coloneqq \fpartial s {s'} A \cdot \fpartial {t'} t A \in \IR$.
    
    Each of the $B_{s', t'}$ has width at most $w-1$ since each layer of $A$ contains at least one node from $V^s$ or $V^t$ and is thus removed in the construction of $B_{s',t'}$.
    We have $|V^s|$, $|V^t|$ choices for $s'$ and $t'$.
    Since $V^s$ does not reach the sink layer and $V^t$ doesn't reach the source layer, we have \footnote{We can even bound $|V^s|, |V^t| \leq d-n+1$ due to the degree of $\imm k n$, but this is not needed for the remaining proofs.} $|V^s|, |V^t| < d$, finishing the proof.
\end{proof}

Using the width reduction, together with the explicit bound on the number of monomials computed by width $2$ mABPs already allows us to separate width $3$ and width $4$ mABPs with an exponential gap on the depth.

\begin{corollary}
    \label{cor:3vs4}
    Any width $3$ mABP computing $\imm 4 n$ requires depth $d \geq 2^{\frac{1}{2}(n-1)}$.
\end{corollary}
\begin{proof}
    We decompose a width $w$ mABP $A$ of depth $d$ for $\imm 4 n$ using \cref{lem:widthreduce} into less than $d^2$ width $2$ mABPs.
    Each of these mABPs can contribute at most $2^{n-1}$ monomials by \cref{lem:depthreduce}.
    Since $\imm 4 n$ has at least $4^{n-1}$ monomials, this requires $d \geq 2^{\frac{1}{2}(n-1)}$.
\end{proof}

\subsection{Densification}

\cref{cor:3vs4} already uses the Decomposition Lemma (\cref{lem:widthreduce}) once to reduce the width of the mABP by one.
However, to further separate more general widths we would like to inductively reduce the width further.
There is just one problem:
After applying \cref{lem:widthreduce}, the resulting mABPs potentially only compute a $\frac{1}{d^2}$ fraction of the monomials of $\imm k n$ each which is not sufficient to invoke the Decomposition Lemma again.
However, we are able to prove a \enquote{Densification Lemma}: If, instead of looking at $\imm k n$, we are content with searching for a \emph{smaller} $\imm k m$ for some $m < n$, we can indeed find a projection resulting in a large enough fraction of monomials of $\imm k m$ to use the Decomposition Lemma once more.

\begin{figure}
    \centering
    \includegraphics[width=0.8\linewidth]{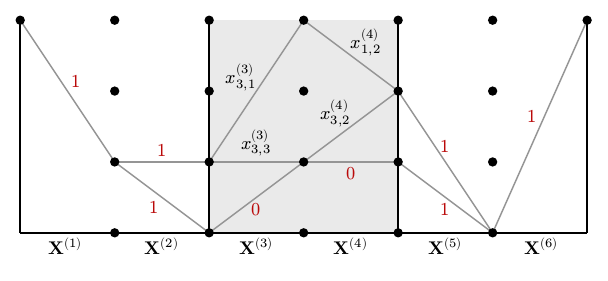}
    \caption{The Densification Lemma, visualized on the corresponding mABP. We split the $n = 6$ matrices into groups of size $m = 2$, then project onto the second group, with $\kappa = 3$ and $\lambda = 2$. The red labels represent those changed by projection $\pi^{(1)}_{3,2}$.}
    \label{fig:densification}
\end{figure}

\begin{lemma}[Densification Lemma]
    \label{lem:make_dense}
    Let $k \geq 1$ and let $f$ be a monotone polynomial with $\supp(f) \subseteq \supp(\imm k n)$ for some $n \in \IN$ and $|\supp(f)| > \alpha \cdot |\supp(\imm k n)| = \alpha \cdot k^{n-1}$ for some $0 < \alpha < 1$.

    Then, for all $m \mid n$, there is a projection $\pi$ with $\supp(\pi(f)) \subseteq \supp(\imm k m)$ such that $|\supp(\pi(f))| > \alpha^{\frac{m}{n}} \cdot |\supp(\imm k m)| = \alpha^{\frac{m}{n}} \cdot k^{m-1}$.
\end{lemma}
\begin{proof}
    We split the matrices $\bX^{(1)}, \ldots, \bX^{(n)}$ into groups of size $m$.
    Define the projections
    \[
        \pi^{(i)}_{\kappa,\lambda}(x^{(j)}_{\mu, \nu}) := \begin{cases}
            1 & \text{if $j \leq i \cdot m$ or $j > (i+1) \cdot m$}\\
            0 & \text{if $j = i \cdot m + 1$ and $\kappa \neq \mu$}\\
            0 & \text{if $j = (i+1) \cdot m$ and $\lambda \neq \nu$}\\
            x^{(j)}_{\mu, \nu} & \text{otherwise}\,.
        \end{cases}
    \]
    If we have $f = \imm k n$, then the projection $\pi^{(i)}_{\kappa,\lambda}$ projects $f$ onto a polynomial with the same support as the $(\kappa, \lambda)$ entry of the product $\bX^{(1+i\cdot m)} \cdots \bX^{(m+i\cdot m)}$.
    
    For general $f$ with $\supp(f) \subseteq \supp(\imm k n)$, every monomial $p$ of $f$ has a unique sequence $\kappa_0, \ldots, \kappa_{\frac{n}{m}}$, such that $\pi^{(i)}_{\kappa_i,\kappa_{i+1}}(p) \neq 0$ for all $i \in \{0, \ldots, \frac{n}{m}-1\}$.
    Furthermore, any two monomials with the same sequence are mapped to distinct monomials under at least one of the projections $\pi^{(i)}_{\kappa_i,\kappa_{i+1}}$.

    For instance, we have that (see also \cref{fig:densification})
    \begin{multline*}
        \pi_{3,2}^{(1)}(x_{1,3}^{(1)} x_{3,3}^{(2)} x_{3,1}^{(3)} x_{1,2}^{(4)} x_{2,4}^{(5)} x_{4,1}^{(6)}
        +
        x_{1,3}^{(1)} x_{3,3}^{(2)} x_{3,3}^{(3)} x_{3,2}^{(4)} x_{2,4}^{(5)} x_{4,1}^{(6)} 
        +
        x_{1,3}^{(1)} x_{3,3}^{(2)} x_{3,3}^{(3)} x_{3,3}^{(4)} x_{3,4}^{(5)} x_{4,1}^{(6)} \\
        + x_{1,3}^{(1)} x_{3,4}^{(2)} x_{4,3}^{(3)} x_{3,2}^{(4)} x_{2,4}^{(5)} x_{4,1}^{(6)} 
        + x_{1,3}^{(1)} x_{3,4}^{(2)} x_{4,3}^{(3)} x_{3,3}^{(4)} x_{3,4}^{(5)} x_{4,1}^{(6)}) 
        = 
        x_{3,1}^{(3)} x_{1,2}^{(4)} + x_{3,3}^{(3)} x_{3,2}^{(4)}.
    \end{multline*}

    Assume for the sake of contradiction that all the projections $\pi^{(i)}_{\kappa,\lambda}$ project $f$ onto polynomials with support size $\leq \alpha^{\frac{m}{n}} \cdot k^{m-1}$.
    Then, for every sequence $\kappa_0, \ldots, \kappa_{\frac{n}{m}}$, there are at most $(\alpha^{\frac{m}{n}} \cdot k^{m-1})^{\frac{n}{m}}$ monomials in $f$ with this specific sequence.
    There are $k^{\frac{n}{m}-1}$ many such sequences (we always have $\kappa_0 = \kappa_{\frac{n}{m}} = 1$), which leads to $f$ having at most $(\alpha^{\frac{m}{n}} \cdot k^{m-1})^{\frac{n}{m}} \cdot k^{\frac{n}{m}-1} = \alpha \cdot k^{n-1}$ monomials, a contradiction.
    
    The desired projection is now any of the $\pi^{(i)}_{\kappa,\lambda}$ projecting $f$ onto a polynomial of support size $>\alpha^{\frac{m}{n}} \cdot k^{m-1}$, composed with the renaming
    \begin{align*}
        x^{(1+i\cdot m)}_{\kappa,\nu} &\mapsto x^{(1)}_{1, \nu}\\
        x^{(m+i\cdot m)}_{\mu,\lambda} &\mapsto x^{(m)}_{\mu, 1}\\
        x^{(j+i\cdot m)}_{\mu,\nu} &\mapsto x^{(j)}_{\mu, \nu} \quad\text{for all $j \in \{2, \ldots, m-1\}$}\,. \qedhere
    \end{align*}
\end{proof}

\subsection{Bringing it All Together}

\begin{theorem}
    \label{thm:main_separation}
    Let $3 \leq w < k$.
    Then, any width $w$ mABP computing a monotone polynomial $f$ with $\supp(f) = \supp(\imm k n)$ requires depth $d \geq \left(\frac{k}{w-1}\right)^{\frac{1}{2}\left\lfloor n^{\sfrac{1}{(w-2)}} \right\rfloor - 1}$.
\end{theorem}
\begin{proof}
    W.l.o.g.\ we only prove this result for $n$ where $N := n^{\sfrac{1}{(w-2)}}$ is an integer.
    If this is not the case we can always reduce $n$ by projecting away some matrices $\bX^{(i)}$ using $x^{(i)}_{1,1} \mapsto 1$ and $x^{(i)}_{\mu,\nu} \mapsto 0$ for $\mu > 1$ or $\nu > 1$.
    Furthermore, we ignore the cases where $N \leq 2$.
    Here, the lower bound of $d \geq 1$ is trivial.

    Then, assume for the sake of contradiction that there is a width $w$ mABP $A$ of depth $d < \left(\frac{k}{w-1}\right)^{\frac{N}{2} - 1}$.

    The proof proceeds in an inductive fashion: 
    Initially we have $n_0 = n$, $\alpha_0 = 1$, $w_0 = w$, $f_0 = f$ and $A_0 = A$.
    Here, $\alpha_i$ denotes the fraction of monomials of $\imm k {n_i}$ that are part of $f_i$, i.e.\ $|\supp(f_i)| \geq \alpha_i |\supp(\imm k {n_i})|$.
    We repeatedly use the Decomposition Lemma (\cref{lem:widthreduce}) to decompose $A_i$ into less than $d^2$ width $w_i-1$ mABPs.
    Using the pigeonhole principle, at least one of these mABPs, let's say $B_i$, has to compute a monotone polynomial $g_i$ with $|\supp(g_i)| > \frac{\alpha_i}{d^2} |\supp(\imm k {n_i})|$.
    Note that, due to the decomposition being monotone, we also have $\supp(g_i) \subseteq \supp(\imm k {n_i})$.
    We then choose an $n_{i+1}$, and, using the Densification Lemma (\cref{lem:make_dense}), from $g_i$, we obtain a monotone polynomial $f_{i+1}$ with $\supp(f_{i+1}) \subseteq \supp(\imm k {n_{i+1}})$ and $|\supp(f_{i+1})| > \alpha_{i+1} \cdot |\supp(\imm k {n_{i+1}})|$ for $\alpha_{i+1} = \left(\frac{\alpha_{i}}{d^2}\right)^{\frac{n_{i+1}}{n_{i}}}$ together with a projection $\pi$ mapping $g_i$ to $f_{i+1}$.
    Applying $\pi$ to $B_i$, we obtain an mABP $A_{i+1}$ of width $w_{i+1} := w_i - 1$ and depth at most $d$, computing $f_{i+1}$.

    This process stops once we reach $w_{w-3} = 3$.
    We use \cref{lem:widthreduce} one last time to decompose $A_{w-3}$ into less than $d^2$ width $2$ mABPs.
    Using \cref{lem:depthreduce}, each of these can contribute at most $2^{n_{w-3}-1}$ monomials, which our choice of $n_1, \ldots, n_{w-3}$ will lead to a contradiction.

    We choose $n_i \coloneqq N^{w-2-i}$, leading to the recursion $\alpha_{i+1} = \left(\frac{\alpha_{i}}{d^2}\right)^{\frac{1}{N}}$.
    We can lower bound each of the $\alpha_i$ for $i > 0$ as follows using the recursion and $\alpha_0 = 1$:
    \begin{align*}
        \ln (\alpha_i) &= \frac{1}{N}(\ln(\alpha_{i-1}) - 2\ln(d))\\
        &\,\,\, \vdots\\
        &= \frac{1}{N^i}\ln(\alpha_0) - 2 \ln(d) \cdot \sum_{j=1}^i\frac{1}{N^j}\\
        &= - 2 \ln(d) \cdot \frac{1-N^{-i}}{N-1}\\
        &> - (N-2) \ln\left(\frac{k}{w-1}\right) \cdot \frac{1-N^{-i}}{N-1}\\
        &\geq - \ln\left(\frac{k}{w-1}\right) = \ln\left(\frac{w-1}{k}\right) \geq \ln\left(\frac{w_i}{k}\right)
    \end{align*}
    Therefore, $\alpha_i > \frac{w_i}{k}$ and hence the condition for \cref{lem:widthreduce} is fulfilled at each step.
    Moreover, the final decomposition of $A_{w-3}$ shows that $f_{w-3}$ has less than $2^{n_{w-3}-1} \cdot d^2 = 2^{N-1} \cdot d^2$ monomials, a contradiction:
    \begin{align*}
        |\supp(f_{w-3})| &< 2^{N-1} d^2
        < 2^{N-1} \left(\frac{k}{w-1}\right)^{N-2}
        \leq 2^{N-1} \left(\frac{k}{2}\right)^{N-2}\\
        &= \frac{2}{k} k^{N-1}
        < \alpha_{w-3} k^{N-1}
        \leq |\supp(f_{w-3})| \qedhere
    \end{align*}
\end{proof}

This lower bound directly yields our desired strict hierarchy, not just for constant-width mABPs, but also for mABPs with slowly growing width.

\begin{corollary}
    \label{cor:main_hierarchy_const}
    Let $1 \leq w < k$.
    Then $\ccMonVBPk{w} \subsetneq \ccMonVBPk{k}$.
\end{corollary}
\begin{proof}
    If $w \geq 3$ this is a direct consequence of \cref{thm:main_separation}.
    For $w = 2$ it follows from \cref{cor:2vs3} and for $w = 1$ it is obvious (an mABP of width $1$ can only compute polynomials that factor into affine linear terms).
\end{proof}

\begin{corollary}
    \label{cor:main_hierarchy_fns}
    $\ccMonVBPk{w(n)} \subsetneq \ccMonVBPk{k(n)}$ for any functions $w, k$ with 
    \begin{equation*}
        \limsup_{n\to\infty} w \cdot \frac{\log\log(n)}{\log(n)} < 1 \quad\text{and}\quad \liminf_{n\to\infty} \frac{k(n)}{w(n)-1} > 1.
    \end{equation*}
\end{corollary}
\begin{proof}
    Consider the family $f_n = \imm {k(n)} n$ which is in $\ccMonVBPk{k(n)}$.
    By \cref{thm:main_separation}, $f_n$ requires width $w(n)$ mABPs of depth at least $d_n \geq \left(\frac{k(n)}{w(n)-1}\right)^{\frac{1}{2}\left\lfloor n^{\sfrac{1}{(w(n)-2)}} \right\rfloor - 1}$.
    Since $\liminf_{n\to\infty} \frac{k(n)}{w(n)-1} > 1$, there is some $\eps_1 > 0$, s.t.\ $\frac{k(n)}{w(n)-1} \geq 1 + \eps_1$ for all large enough $n$.
    Similarly, due to $\limsup_{n\to\infty} w \cdot \frac{\log\log(n)}{\log(n)} < 1$, there is some $\eps_2 > 0$, s.t.\ $w(n)-2 \leq (1-\eps_2) \cdot \frac{\log(n)}{\log\log(n)}$ for all large enough $n$.
    We obtain an exponent of
    \begin{align*}
        \frac{1}{2}\left\lfloor n^{\frac{1}{w(n)-2}} \right\rfloor - 1 = \frac{1}{2}\left\lfloor n^{\frac{\log\log(n)}{(1-\eps_2)\log(n)}} \right\rfloor - 1 = \frac{1}{2}\left\lfloor \log(n)^{\frac{1}{(1-\eps_2)}} \right\rfloor - 1 = \omega(\log(n))\,.
    \end{align*}
    Thus, $d_n \geq (1+\eps_1)^{\omega(\log(n))} = n^{\omega(1)}$ and $(f_n) \notin \ccMonVBPk{w(n)}$.
\end{proof}

Note that the conditions in \cref{cor:main_hierarchy_fns} are implied by $w \in o\left(\sfrac{\log(n)}{\log\log(n)}\right)$ and $w \in o(k(n))$ respectively.
However, \cref{cor:main_hierarchy_fns} for example even works for $w(n) = \sfrac{\log(n)}{2\log\log(n)}$ and $k(n) = \sfrac{\log(n)}{\log\log(n)}$.

\section{Connections to Constant-Depth Monotone Formulas}

A hierarchy similar to the constant-width hierarchy established in \cref{cor:main_hierarchy_const} has existed for monotone formulas of constant depth, see \cite{DBLP:conf/stoc/KlawePPY84}, \cite{DBLP:conf/focs/ChillaraEL018}. 
This begs the question: How are the two hierarchies related?

As a reminder, throughout this section, the class $\ccMonVFk{d}$ denotes $p$-families computable by un\-bounded-fan-in monotone formulas of depth $\leq d$ and polynomial size.

We start by showing a construction that can convert monotone formulas of unbounded fan-in with addition depth $\dadd$ to mABPs of width $2\dadd + 1$ with polynomial size.
Simulating multiplication gates is the easiest and well known:
\begin{figure}
    \centering
    \includegraphics[width=0.85\linewidth]{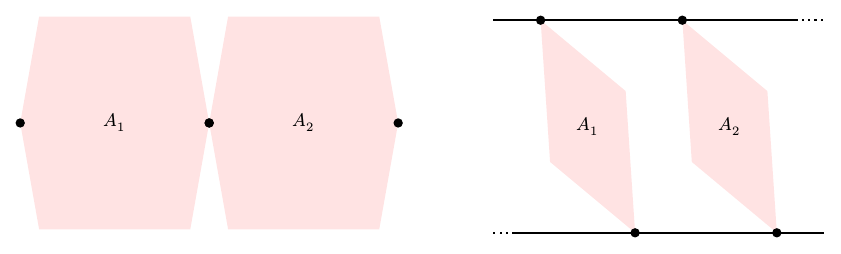}
    \caption{The construction of \cref{lem:abpprod} (on the left) and \cref{lem:abpsum}  (on the right).}
    \label{fig:series-parallel}
\end{figure}
\begin{lemma}
    \label{lem:abpprod}
    Let $A_1, \ldots, A_r$ be mABPs of depths $d_1, \ldots, d_r$ and widths $w_1, \ldots, w_r$, then there is an mABP $A$ of depth $\sum_{i=1}^r d_i$ and width $\max \{w_1, \ldots, w_r\}$ computing $\prod_{i=1}^r \ffull{A_i}$.
\end{lemma}
\begin{proof}
    Identifying the sink of $A_i$ with the source of $A_{i+1}$ for $1 \leq i < r$ directly constructs $A$ with the desired properties.
\end{proof}

Simulating addition gates with mABPs is different to the usual construction of ABPs from formulas since we want to only use width sparingly.
In particular, we use the structure of long constant paths we already observed in our lower bound in \cref{lem:widthreduce}.
\begin{lemma}
    \label{lem:abpsum}
    Let $A_1, \ldots, A_r$ be mABPs of depths $d_1, \ldots, d_r$ and widths $w_1, \ldots, w_r$, then there is an mABP $A$ of depth $\sum_{i=1}^r d_i$ and width $2 + \max \{w_1, \ldots, w_r\}$ computing $\sum_{i=1}^r \ffull{A_i}$.
\end{lemma}
\begin{proof}
    We start constructing $A$ from a path $s_0, \ldots, s_{\ell_1}$ from layer $0$ to layer $\ell_1 = \sum_{i=1}^{r-1}d_i$ and a path $t_{d_1}, \ldots, t_{\ell_2}$ from layer $d_1$ to layer $\ell_2 = \sum_{i=1}^{r}d_i$.
    All edges along these two paths have weight $1$.
    For $1 \leq j \leq r$ we now place $A_j$ in such a way in $A$ that its source is identified with $s_{\sum_{i=1}^{j-1}d_i}$ and its sink is identified with $t_{\sum_{i=1}^jd_i}$.
    This construction is visualized in \cref{fig:series-parallel}.

    We observe that $A$ is indeed a correctly formed mABP as each of the $A_j$ has depth $d_j$ and its source and sink are identified with $s_{\sum_{i=1}^{j-1}d_i}$ and $t_{\sum_{i=1}^jd_i}$ which are in layers that are $d_j$ apart.
    The width of $A$ is at most $2 + \max \{w_1, \ldots, w_r\}$ due to the $A_j$ being in disjoint layers except their sources and sinks.
    $A$ clearly computes $\sum_{i=1}^r f_{A_i}$ since each path from $s_0$ to $t_{\ell_2}$ crosses exactly one of the $A_j$ from source to sink.
\end{proof}

Using \cref{lem:abpprod,lem:abpsum} recursively on a monotone formula, we obtain:
\begin{corollary}
    \label{cor:formulatoabp}
    Let $f$ be computed by a monotone formula with addition depth $\dadd$ and size $s$.
    Then, there is an mABP of width $2\dadd + 1$ and depth $s$ computing $f$.
\end{corollary}

This gives a conversion from bounded-depth monotone formulas to mABPs.
To convert in the opposite direction, we use a straightforward generalization of the \enquote{standard binary splitting method} (we use the name as in \cite{DBLP:conf/icalp/Snir80}) where the matrices are split into more than two blocks.
It is well known (see \cite{DBLP:conf/stoc/TavenasL022}, albeit only called out for product depth and not specifically monotone), but we give a proof for completeness.
\begin{lemma}
    \label{lem:imm_to_depth_formula}
    The polynomial $\imm k n$ for $k \geq 2$ can be computed by arbitrary-fan-in monotone formulas of addition depth $\dadd$ with $n \cdot k^{\dadd(\lceil n^{\sfrac{1}{\dadd}} \rceil-1)}$ leaves for every $\dadd \geq 1$.
    Moreover, these formulas have an addition gate as their root gate.
\end{lemma}
\begin{proof}
    Let $k \geq 2$ be fixed.
    We construct the family of polynomials
    \begin{equation*}
        f_{\mu,\nu,N,\delta,i} = (\bX^{(i)} \cdot \bX^{(i+1)} \cdots \bX^{(i+N^\delta-1)})_{\mu, \nu},
    \end{equation*} where, as usual, the $\bX^{(j)}$ are $k \times k$ matrices of variables.
    Polynomials of the form $\imm k {N^{\dadd}}$ can be directly retrieved as $f_{1,1,N,\dadd,1}$ while for general $\imm k n$ we construct $\imm k {N^{\dadd}}$ for $N = \lceil n^{\sfrac{1}{\dadd}} \rceil$ and project away the surplus matrices.

    For $\delta = 0$, we have $f_{\mu,\nu,N,\delta,i} = x^{(i)}_{\mu, \nu}$, which can be represented by a formula of addition depth $0$ and has size $1$.
    Note that, for this special case, the root node is not an addition gate.
    For $\delta \geq 1$, we decompose 
    \begin{align}
        f_{\mu,\nu,N,\delta,i} = \sum_{\lambda_0, \ldots, \lambda_N}\prod_{j=1}^{N}f_{\lambda_{j-1}, \lambda_j, N, \delta-1,i+(j-1) \cdot N^{\delta-1}} \label{eq:imm_to_depth_formula:recursion}
    \end{align}
    where $\lambda_0 = \mu, \lambda_N = \nu$ and all other $\lambda_j \in \{1, \ldots, k\}$.
    By induction, this decomposition produces arbitrary-fan-in monotone formulas of addition depth $\delta$ and $s(k, N, \delta) = k^{N-1} \cdot N \cdot s(k, N, \delta-1)$ leaves, which solves to $N^\delta k^{\delta(N-1)}$.

    So far, the size would obtain a factor of $\lceil n^{\sfrac{1}{\dadd}} \rceil^{\dadd}$ at the front.
    This is caused by first constructing $\imm k {N^{\dadd}}$ and then projecting away the surplus matrices.
    By choosing the recursive sub-polynomials in \cref{eq:imm_to_depth_formula:recursion} non-uniformly, we can avoid constructing these surplus matrices altogether, turning the front factor to just $n$.
\end{proof}
For addition depth $\lceil\log(n)\rceil$ this recovers the classical bound of $(2k)^{\lceil \log(n) \rceil}$ leaves for $\imm k n$.
Generalizing the argument of \textcite{DBLP:conf/icalp/Snir80} that proves a lower bound of $(2k)^{\log(n)}$ leaves in any monotone formula, we can even prove that the construction of \cref{lem:imm_to_depth_formula} produces the minimum number of leaves for a fixed depth, up to the rounding of the $n^{\sfrac{1}{\dadd}}$ terms:

\begin{theorem}
    \label{thm:depth_snir_imm}
    Let $F$ be an arbitrary-fan-in monotone formula of multiplication depth $\dmult$ for some $\imm k n$.
    Then $F$ has at least $n \cdot k^{\dmult(n^{\sfrac{1}{\dmult}} - 1)}$ leaves.
\end{theorem}
We postpone the \hyperlink{proof:depth_snir_imm}{proof} to \cref{sec:snir}.

Note that for $\imm k n$ with $n,k \geq 2$, addition and multiplication depth of monotone formulas are equal.
The root gate has to always be an addition gate due to the irreducibility of $\imm k n$ for $n, k \geq 2$, while the gates right before the leaves cannot be addition gates with an argument similar to \cref{lem:edges_const_or_single_var}. Thus, the bounds in \cref{thm:depth_snir_imm} and \cref{lem:imm_to_depth_formula} do match (up to rounding).

As a consequence, we recover a result by \textcite{DBLP:journals/siamcomp/ChillaraL019}.

\begin{corollary}
    \label{cor:imm2_not_in_constant_depth_formulas}
    $\ccMonVBPk{2} \not\subseteq \ccMonVFk{O(1)}$
\end{corollary}
\begin{proof}
    Consider the $p$-family $(\imm 2 n) \in \ccMonVBPk{2}$.
    If $(\imm 2 n) \in \ccMonVFk{O(1)}$, then there is some $\dmult \in \IN$, such that $\imm 2 n$ has monotone formulas with multiplication depth $\dmult$ and size $s(n) \in \poly(n)$.
    However, by \cref{thm:depth_snir_imm}, monotone formulas with multiplication depth $\dmult$ for $\imm 2 n$ require size $\Omega\left(n2^{\dmult(n^{\sfrac{1}{\dmult}}-1)}\right)$ which is superpolynomial in $n$.
\end{proof}
Together with \cref{cor:formulatoabp}, this gives us that constant-width mABPs as a whole are more powerful than constant-depth monotone formulas.
\begin{corollary}
    \label{lem:VBPkO1_vs_VFkO1}
    $\ccMonVFk{O(1)} \subsetneq \ccMonVBPk{O(1)}$.
\end{corollary}

As another consequence of \cref{cor:imm2_not_in_constant_depth_formulas}, using the fact that any depth $d$ formula with arbitrary fan-in w.l.o.g.\ has addition depth $\dadd \leq \lceil \frac{d}{2} \rceil $, we get that the inclusion from \cref{cor:formulatoabp} is strict:
\begin{corollary}
    \label{cor:formula_to_almost_same_width_abp}
    For every $k \geq 1$, $\ccMonVFk{k} \subsetneq \ccMonVBPk{2 \lceil\frac{k}{2}\rceil + 1}$.
\end{corollary}

\cref{cor:formula_to_almost_same_width_abp} tells us that, given a fixed depth $k$, for a constant-width mABP to be able to compute it, roughly the same width is enough. We can further ask: Is that the lowest width that can reproduce it? The following result tells us that roughly $\frac{k}{2}$ is definitely \emph{not} enough.
\begin{theorem}
    \label{thm:ABPkNotInFormula2k}
    For every $k \geq 2$, $\ccMonVFk{2k} \not\subseteq \ccMonVBPk{k-1}$.
\end{theorem}
\begin{proof}
    Consider the $p$-family $f_t = \imm k {\lfloor\log(t)\rfloor^k}$.
    Using \cref{lem:imm_to_depth_formula}, we see that $f_t$ has monotone formulas of addition depth $k$ of size $O((\log t)^k k^{k \log(t)}) = O((\log t)^k t^{k \log k})$, i.e.\ of polynomial size.
    The roots of each of these formulas is an addition gate, so the total depth is at most $2k$.
    Consequently, $(f_t) \in \ccMonVFk{2k}$.

    For $k=2,3$ we directly see $(f_t) \notin \ccMonVBPk{k-1}$: width $1$ or $2$ mABPs are too weak to compute the required iterated matrix multiplications at all (see \cref{cor:2vs3} for the width $2$ case).
    For $k \geq 4$, we can apply \cref{thm:main_separation} to obtain that any width $k-1$ mABP for $f_t$ requires depth at least
    $
        \left(\frac{k}{k-2}\right)^{\frac{1}{2}\left\lfloor (\lfloor\log(t)\rfloor^k)^{\frac{1}{k-3}} \right\rfloor - 1} \in t^{\Omega\left(\log(t)^{\frac{3}{k-3}}\right)},
    $
    a quasi-polynomial lower bound implying $(f_t) \notin \ccMonVBPk{k-1}$.
\end{proof}

\section{Separations in the Superconstant Regime}
\label{sec:superconstant_separations}

We will now pivot our attention to the superconstant regime, focusing on monotone formulas first and then on comparing mABPs of different growing widths.

\subsection{Monotone Formulas}
\label{sec:placement_of_mVF}

It is well known that $\ccMonVPe \subsetneq \ccMonVBP$ \cite{DBLP:conf/icalp/Snir80}.
In this section we will tighten this separation more and see where $\ccMonVPe$ sits in the mABP hierarchy.

In a recent result, \textcite{DBLP:conf/coco/FournierLM0T23} have shown that monotone formulas allow for a depth reduction:
\begin{lemma}[Theorem 1 in \cite{DBLP:conf/coco/FournierLM0T23}]
    \label{lem:formuladepthreduce}
    Let $F$ be a homogeneous unbounded-fan-in algebraic formula of size $s$ computing a polynomial $f$ of degree $d \geq 2$.
    Then $f$ is also computed by a homogeneous unbounded-fan-in formula $F'$ of size $\poly(s)$ and depth $O(\log d)$.
    Moreover, if $F$ is monotone and/or non-commutative, then so is $F'$.
\end{lemma}

Combining \cref{cor:formulatoabp,lem:formuladepthreduce} we already see that homogeneous algebraic formulas of polynomial size can be converted to mABPs of logarithmic width.
With a standard homogenization trick, we can extend this also to inhomogeneous algebraic formulas:

\begin{proposition}
\label{prop:mVF_in_mVBP_O_logn}
    $\ccMonVPe \subsetneq \ccMonVBPk{O(\log(n))}$
\end{proposition}
\begin{proof}
    Let $f$ be a polynomial of degree $d$, computed by some monotone algebraic formula $F$ of size $s$.
    Introduce a fresh variable $y$ and structurally use $y$ to homogenize $F$.
    Inductively, starting at the leaves, whenever we encounter an addition gate where the children compute (homogeneous) polynomials of different degrees, multiply the children of smaller degree with appropriate powers of $y$ to reach the same degree as the child with the highest degree.
    The resulting polynomial $\tilde{f}$ is now homogeneous of degree $d$, computed by a monotone algebraic formula $\tilde{F}$ of size $O(s \cdot d)$.
    Note that here we used the fact that for monotone formulas, the formal degree and the degree coincide and thus the powers of $y$ cannot increase the degree. 

    Starting with a $p$-family $(f_n)$ with monotone formulas of size $\poly(n)$ and degree $\poly(n)$ and applying \cref{cor:formulatoabp,lem:formuladepthreduce} we obtain a family of mABPs of width $O(\log(n))$ computing $\tilde{f}_n$.
    Finally, the substitution $y \mapsto 1$ yields mABPs of width $O(\log(n))$ and with polynomial size computing $f_n$ itself.

    To prove strictness, we use the well-known result by \textcite{DBLP:conf/icalp/Snir80} that $\imm k n$ requires monotone formulas of size\footnote{Technically, this is a lower bound on the number of leaves of such a formula, but it is also a lower bound on the size itself.} at least $(2k)^{\log(n)}$.
    In particular, $\imm {\lfloor\log(n)\rfloor} n$ requires monotone formulas of size $\Omega(n^{\log(2\log(n))})$ while obviously having width $\lfloor\log(n)\rfloor$ mABPs of depth $n$.
\end{proof}
Indeed, using the same argument, for any unbounded function $k(n)$, $\imm {k(n)} n$ requires formulas of size $\Omega(n^{\log(2 k(n))})$. Thus,
\begin{corollary}
\label{cor:mVBP_k_not_in_mVF_if_unbounded}
    $\ccMonVBPk{k(n)} \not\subseteq \ccMonVPe$, unless $k$ is bounded. 
\end{corollary}
For bounded $k$, however, things are different. Here, it is well known that $\ccMonVBPk{k} \subseteq \ccMonVF{}$ by the standard binary splitting method. The following theorem shows that this inclusion is strict.

\begin{theorem}
\label{thm:poly_in_mVF_but_not_in_slow_growing_width_mVBP}
    There is a $p$-family $(f_t)$ such that $(f_t) \in \ccMonVPe$, but for every $w(t) \in o\left(\frac{\log t}{(\log \log t)^2}\right)$, $f_t \notin \ccMonVBPk{w(t)}$.
\end{theorem}
\begin{proof}
    Our polynomial family is given by $f_t \coloneqq \imm{k_t}{n_t}$ for
    \begin{align*}
        k_t \coloneqq \left\lfloor \frac{\log t}{2} \right\rfloor \qquad n_t \coloneqq \left\lfloor t^{\frac{1}{\log \log t}} \right\rfloor.
    \end{align*}

    We will first show the upper bound using the usual binary splitting method to construct a monotone formula for $f_t$ of polynomial size. 
    The number of leaves of the formula is given recursively by $L_{k_t}(n_t) = 2 k_t \cdot L_{k_t}(\lceil\frac{n_t}{2}\rceil)$, which solves to $L_{k_t} = (2k_t)^{\lceil\log n_t \rceil}.$ Since
    \begin{equation*}
        L_{k_t}(n_t) \leq (2k_t)^{\log n_t} \cdot 2k_t \leq \log t^{\log n_t} \cdot 2k_t \leq t \cdot 2k_t \in \poly(t),
    \end{equation*}
    and the size of the formula is at most twice the number of leaves, we conclude that $(f_t) \in \ccMonVF$.

    Now we prove the lower bound. W.l.o.g.\ assume $w(t) \geq 3$. Applying \cref{thm:main_separation}, we obtain the following lower bound for the depth $d_t$ of a width $w(t)$-mABP computing $f_t$:
    \begin{equation*}
        d_t \geq \left(\frac{k_t}{w(t)-1} \right)^{\frac{1}{2}\lfloor n_t^{\frac{1}{w(t)-2}} \rfloor - 1}
    \end{equation*}
    Note that $\frac{k_t}{w(t)-1} \geq 2$ for large enough $t$ since $w \in o\left(\log t\right)$. Additionally, due to $w \in o\left(\frac{\log t}{(\log \log t)^2}\right)$, we have that $n_t^{\frac{1}{w(t)-2}} \in \omega(\log t)$. Therefore,
    \begin{equation*}
        d_t \geq 2^{\frac{1}{2} \lfloor \omega(\log t) \rfloor - 1},
    \end{equation*}
    which is superpolynomial. Hence, $(f_t) \notin \ccMonVBPk{w(t)}$.
    
\end{proof}

In particular, by setting $w(t)$ to a constant, we obtain the following lower bound:
\begin{corollary}
\label{cor:constant_width_strict_subset_of_mVF}
    $\ccMonVBPk{w} \subsetneq \ccMonVF$ for any $w\in\mathbb{N}$.
\end{corollary}

This is in contrast to the non-monotone world, in which we know that $\ccVPe = \ccVBPk{3}$. The results paint an interesting picture of the position of $\ccMonVF$ compared to mABPs of width $O(k(n))$: If $k(n)$ is constant, they lie strictly below $\ccMonVF$, if it is slow-growing (in $o\left(\frac{\log n}{(\log \log n)^2}\right)$), they are incomparable to $\ccMonVF$ and if $k(n) \in \Omega(\log n)$, they are strictly above $\ccMonVF$ (see \cref{fig:hierarchy}).

\subsection{Coarse Separations for mABPs}
\label{sec:separations_superconstant_mABPs}

While not using \cref{thm:main_separation}, we can get some coarse separations of mABPs of larger widths in the range $\Omega(\sfrac{\log(n)}{\log\log(n)})$.
We achieve this by a detour to monotone formulas for $\imm k n$, using the result by \cite{DBLP:conf/icalp/Snir80} that the number of leaves in any monotone formula computing $\imm k n$ is at least $(2k)^{\log(n)}$, combined with the upper bound that this is tight up to rounding the exponent. 

\begin{lemma}
    \label{lem:logvslin}
    $\ccMonVBPk{O(\log(n))} \subsetneq \ccMonVBP$
\end{lemma}
\begin{proof}
    Consider the $p$-family $(\imm n n) \in \ccMonVBP$.
    Now assume for the sake of contradiction, that $(\imm n n) \in \ccMonVBPk{O(\log(n))}$, i.e.\ there are mABPs of width $w(n)$ and depth $d(n)$ computing $\imm n n$ with $w \in O(\log(n))$ and $d \in \poly(n)$.
    By \cref{lem:simplicity}, we can assume all these mABPs to be simple.
    In particular, there are projections $\pi_n$, such that $\pi_n(\imm {w(n)} {d(n)}) = \imm n n$.

    Using the usual splitting construction, there are monotone formulas with $(2w(n))^{\lceil\log(d(n))\rceil}\allowbreak = n^{O(\log\log(n))}$ leaves computing $\imm {w(n)} {d(n)}$.
    Applying $\pi_n$ to this monotone formula, gives monotone formulas $n^{O(\log\log(n))}$ with leaves for $\imm n n$.
    However, $\imm n n$, by \cite{DBLP:conf/icalp/Snir80}, does not allow for monotone formulas with less than $(2n)^{\log(n)} = n^{\Omega(\log(n))}$ leaves.
\end{proof}

Using the same technique, we also get a more general separation:
\begin{proposition}
    \label{cor:coarse_abp_separation_logf_vs_logg}
    Let $w, k$ be unbounded growing functions with $\log w(n) \in o(\log k(n))$.
    Then $\ccMonVBPk{w(n)} \subsetneq \ccMonVBPk{k(n)}$.
\end{proposition}
\begin{proof}
    Similarly to \cref{lem:logvslin}, consider the $p$-family $(\imm {k(n)} n)$.
    Now assume for the sake of contradiction, that there are mABPs of width $w(n)$ and depth $d(n)$ computing $\imm {k(n)} n$ with $d \in \poly(n)$.
    By \cref{lem:simplicity}, we can assume all these mABPs to be simple.
    In particular, there are projections $\pi_n$, such that $\pi_n(\imm {w(n)} {d(n)}) = \imm {k(n)} n$.

    As previously, we can construct monotone formulas with $(2w(n))^{\lceil\log(d(n))\rceil} = n^{O(\log(w(n)))} = n^{o(\log(k(n)))}$ leaves for $\imm {k(n)} n$, while \cite{DBLP:conf/icalp/Snir80} gives a lower bound of $(2k(n))^{\log(n)} = n^{\Omega(\log(k(n)))}$ leaves, again a contradiction.
\end{proof}

\section{Generalizing Snir's Method}
\label{sec:snir}
This section is dedicated to the monotone formula lower bound framework of \textcite{DBLP:conf/icalp/Snir80} and generalizing it to account for bounded-depth monotone formulas, with the goal of proving \cref{thm:depth_snir_imm}.
Note that throughout this section we will often use $d$ instead of $\dmult$ to denote multiplication depth to keep the intermediate formulas readable.

The generalized framework consists of multiple components:
\begin{itemize}
    \item A fixed polynomial $p$ we want to analyze the complexity of.
    
    \item A weight function $w(q)$, denoting some sub-additive and sub-multiplicative measure for polynomials $q$ that is bounded above by $1$ for individual variables and constants.
    Often, $w(q) = |\supp(q)|$ and this is indeed also the setting we will require for the iterated matrix multiplication.
    
    \item A growth function $\growth(n)$ that gives an upper bound on the weight $w(q)$ for any polynomials $q$ of degree $n$.
    It needs to be monotonically non-decreasing, bounded below by $1$, and we require $\growth(n) = w(p)$ if $p$ contains any monomial of degree $n$.
    If all these properties hold, we say that $p$, and the functions $w$ and $\growth$ are compatible.
    In the setting of $w(q) = |\supp(q)|$, $\growth(n)$ gives an upper bound on the number of \enquote{relevant} monomials that can be computed simultaneously at any node of a monotone formula of degree $n$.

    \item A dynamic programming function\footnote{
        This is where we primarily deviate from \textcite{DBLP:conf/icalp/Snir80}, his dynamic programming function was defined without the parameter for depth.}
    $W(n, s, d)$ giving an upper bound on the weight of relevant polynomials $q$ of degree $n$ computed by monotone formulas with $s$ leaves and multiplication depth $d$.
    In particular, this dynamic programming function caps the weight at $\growth(n)$ at every sub-formula, which is the main improvement over direct \enquote{monomial counting} techniques:
    While normal monomial counting tries to count the maximum number of monomials any formula can produce, many such monomials (or combinations of monomials) will ultimately not help in the computation of $p$.
    Even worse, since our computation is monotone, we cannot cancel them, and they will make the computation of $p$ impossible.
    Capping the number of monomials in each sub-formula by its theoretical maximum thus vastly improves the resulting lower bound by giving a more realistic upper bound on the true weight that can be computed in our formula.

    \item A lower bound function $L(n, d)$ which lower bounds the point $s$ where $W(n, s, d) \geq w(p)$, i.e.\ it gives a lower bound on the smallest possible formula that is able to generate enough \enquote{useful} weight to be able to compute $p$.
    We will give a sufficient characterization for $L$ via some recursive inequalities in \cref{thm:snir_general}.
\end{itemize}

In the end, to prove \cref{thm:depth_snir_imm} itself we will only need to find explicit formulas for a weight function $w(q)$, a compatible growth function $\growth(n)$ and a lower bound function $L(n, \dmult)$.
The dynamic programming function $W(n, s, d)$ is only an internal tool in the proof of the framework.

We say a polynomial $q$ is \emph{relevant} for computing $p$ if there are monotone polynomials $a$ and $b$, s.t.\ $p = q \cdot a + b$.
If $p$ is computed by some monotone formula $F$, there cannot be a sub-formula $G$ of $F$ that computes an \enquote{irrelevant} polynomial due to monotonicity.

The next goal is to define a capped upper bound on the weight using the sub-additivity and sub-multiplicativity of the weight, together with the upper bound of the growth function on the weight.

\begin{lemma}
    \label{lem:snir_dp}
    Let $p$ be a monotone polynomial with compatible weight function $w$ and growth function $\growth$.
    Moreover, let $W(n, s, d)$ be the inductively defined function
    \begin{align*}
        W(n, s, d) = \min(
            \growth(n),
            \max(
                &\max_{s = \sum_{i=1}^\ell s_i}(\sum_{i=1}^\ell {W(n, s_i, d})),\\
                &\max_{\substack{n = \sum_{i=1}^\ell n_i\\s = \sum_{i=1}^\ell s_i}}(\prod_{i=1}^\ell {W(n_i, s_i, d-1}))
            )
        )
    \end{align*}
    with base case $W(n, 1, d) = 1$ if $n \in \{0,1\}$ and $d \geq 0$ and $-\infty$ otherwise.
    All decompositions as part of the maxima are over all values of $\ell \geq 2$.

    Then every polynomial $q$ that is relevant for $p$ and computed by some monotone formula with $s$ leaves and multiplication depth at most $\dmult$, has $w(q) \leq W(\deg(q), s, \dmult)$.
\end{lemma}
\begin{proof}
    W.l.o.g.\ we assume the monotone formula to not have any gates of arity $1$.
    Any such gates can always be contracted.
    
    We prove the statement of the lemma via strong induction on $s$.
    For $s=1$, the monotone formula $F$ has a single leaf and hence computes a polynomial $q$ that is either a single variable or a constant and thus $w(q) \leq 1$, matching the base case of $W$.
    
    Otherwise, for $s > 1$, we either have $F = \sum_{i=1}^\ell F_i$ or $F = \prod_{i=1}^\ell F_i$ for some $\ell \geq 2$ and sub-formulas $F_1, \ldots, F_\ell$ with $s_1, \ldots, s_\ell$ leaves and computing polynomials $q_1, \ldots, q_\ell$ of degree $n_1, \ldots, n_\ell$ each.
    Due to the disjointness of the sub-formulas we have $s = \sum_{i=1}^\ell s_i$.
    
    We start with the addition case, i.e.\ $F = \sum_{i=1}^\ell F_i$.
    If $F$ has multiplication depth at most $d$, then the $F_i$ still have multiplication depth at most $d$.
    Additionally, due to monotonicity, we have $n := \deg(q) = \max(n_1, \ldots, n_\ell)$.
    Clearly, the $q_i$ are also relevant for $p$, and thus we can apply the inductive claim and obtain $w(q_i) \leq W(n_i, s_i, d)$.
    Finally, $q$ itself is relevant for $p$ and thus $w(q) \leq \growth(n)$.
    Combined, we have 
    \begin{align*}
        w(q)
            &\leq \min\left(\growth(n),\; \sum_{i=1}^\ell w(q_i)\right)\\
            &\leq \min\left(\growth(n),\; \sum_{i=1}^\ell W(n_i, s_i, d)\right)\\
            &\leq \min\left(\growth(n),\; \sum_{i=1}^\ell W(n, s_i, d)\right)
            \leq W(n, s, d)\,.
    \end{align*}
    Note that $W(n_i, s_i, d) \leq W(n, s_i, d)$ follows directly from $\growth$ being monotonically non-\-de\-crea\-sing, making $W$ non-decreasing in the first parameter.

    Remains the multiplication case, i.e.\ $F = \prod_{i=1}^\ell F_i$.
    If $F$ has multiplication depth at most $d$, then the $F_i$ have multiplication depth at most $d-1$.
    Additionally, we have $n := \deg(q) = \sum_{i=1}^\ell n_i$.
    Again, the $q_i$ are relevant for $p$, and thus we can apply the inductive claim and obtain $w(q_i) \leq W(n_i, s_i, d-1)$.
    Combined with $w(q) \leq \growth(n)$, we have 
    \begin{align*}
        w(q)
            &\leq \min(\growth(n), \prod_{i=1}^\ell w(q_i))\\
            &\leq \min(\growth(n), \prod_{i=1}^\ell W(n_i, s_i, d-1))
            \leq W(n, s, d)\,.
    \end{align*}
\end{proof}

We are now interested in the smallest $s$, s.t.\ $W(\deg(p), s, d) \geq w(p)$, i.e.\ the smallest size where a formula could theoretically have accumulated enough weight to compute $p$.
Since $w(p) = \growth(\deg(p))$ and $W(\deg(p), s, d) \leq \growth(\deg(p))$, this is equivalent to searching for the smallest $s$ with $W(\deg(p), s, d) = \growth(\deg(p))$.
Generalizing a bit, our plan is to find a lower bound $L(n, d)$ for the saturation threshold $t(n, d) := \min \{s \mid W(n, s, d) = \growth(n)\}$.

To find conditions on $L$, we start with the observation that multiplication gates yield more weight for their size than additions.
Moreover, intuitively, the effect of multiplications are maximized whenever we multiply polynomials of maximal weight.
However, when multiplying polynomials of degrees $n_1, \ldots, n_\ell$, the resulting weight cannot exceed $\prod_{i=1}^\ell \growth(n_i)$.
To reach the maximal weight of $\growth(\sum_{i=1}^\ell n_i)$ again, we thus require the use of $\frac{\growth(\sum_{i=1}^\ell n_i)}{\prod_{i=1}^\ell \growth(n_i)}$ additions.
In particular, if $L(n_i, d-1)$ lower bounds the number of leaves required to maximize the weight of a degree $n_i$ polynomial, then the lower bound to maximize the weight of a degree $n = \sum_{i=1}^\ell n_i$ polynomial should not exceed $\frac{\growth(\sum_{i=1}^\ell n_i)}{\prod_{i=1}^\ell \growth(n_i)} \sum_{i=1}^\ell L(n_i, d-1)$.
Note that, since we don't know whether this decomposition of the degree was optimal, we require this condition for any such decomposition.
It turns out, this intuitive condition is also sufficient, when combined with a sensible base case for degree $1$ polynomials.

\begin{theorem}
    \label{thm:snir_general}
    Let $p$ be a monotone polynomial of degree $n$ with compatible weight function $w$ and growth function $\growth$ and let $L$ be a function fulfilling the following inequalities:
    \begin{align}
        L(1, d) &\leq \growth(1) && \text{for all $d \geq 0$}\label{eq:L_base}\\
        L\left(\sum_{i=1}^\ell n_i, d\right) &\leq \frac{\growth(\sum_{i=1}^\ell n_i)}{\prod_{i=1}^\ell \growth(n_i)} \cdot \sum_{i=1}^\ell L(n_i, d-1)&& \text{for all $d, \ell, n_i \geq 1$} \label{eq:L_ind}
    \end{align}
    Then, any arbitrary-fan-in monotone formula of multiplication depth $\dmult$ that computes $p$ requires at least $L(n, \dmult)$ many leaves.
    Note that we allow $L(n, 0)$ to be infinite if there are no multiplication depth $0$ formulas that compute $p$.
\end{theorem}
\begin{proof}
    We claim that $s \geq L(n, d) \cdot \frac{W(n, s, d)}{\growth(n)}$ for all $s \geq 1$ and $d \geq 0$ for the function $W$ from \cref{lem:snir_dp}.
    The statement of the theorem then follows:
    Let $F$ be a monotone formula for $p$ with $s$ leaves and multiplication depth $d$.
    Using \cref{lem:snir_dp}, we have $W(n, s, d) \geq w(p) = \growth(n)$ (by definition of $\growth$).
    Finally, combined with $W(n, s, d) \leq \growth(n)$ we get the lower bound of $s \geq L(n, d) \cdot \frac{W(n, s, d)}{\growth(n)} = L(n, d)$ on the number of leaves.

    We prove the claim via strong induction on $(n, s, d)$ in lexicographic order.
    For $n = 1$, $W(1, s, d) \leq s$ and $L(1, d) \leq \growth(1)$ and thus $s \geq L(1, d) \cdot \frac{W(1, s, d)}{\growth(n)}$.
    For $n > 1$, we distinguish two cases (from the definition of $W$):
    
    If $W(n, s, d) \leq \sum_{i=1}^\ell W(n, s_i, d)$ for some $\ell \geq 2$ and $s = \sum_{i=1}^\ell s_i$, then, by induction, $s_i \geq L(n, d) \cdot \frac{W(n, s_i, d)}{\growth(n)}$, yielding 
    \begin{align*}
        s = \sum_{i=1}^\ell s_i \geq L(n, d) \cdot \frac{\sum_{i=1}^\ell W(n, s_i, d)}{\growth(n)} \geq L(n, d) \cdot \frac{W(n, s, d)}{\growth(n)}\,.
    \end{align*}

    If $W(n, s, d) \leq \prod_{i=1}^\ell W(n_i, s_i, d-1)$ for some $\ell \geq 2$, $n = \sum_{i=1}^\ell n_i$ and $s = \sum_{i=1}^\ell s_i$, then, by induction $s_i \geq L(n_i, d-1) \cdot \frac{W(n_i, s_i, d-1)}{\growth(n_i)}$, yielding
    \begin{align*}
        L(n, d) &\leq \frac{\growth(n)}{\prod_{i=1}^\ell \growth(n_i)} \cdot \sum_{i=1}^\ell L(n_i, d-1)\\
            &\leq \frac{\growth(n)}{\prod_{i=1}^\ell \growth(n_i)} \cdot \sum_{i=1}^\ell s_i \cdot \frac{\growth(n_i)}{W(n_i, s_i, d-1)}\\
            &= \frac{\growth(n)}{\prod_{i=1}^\ell \growth(n_i)} \cdot \frac{ \sum_{i=1}^\ell s_i \cdot \growth(n_i) \cdot \prod_{j\neq i} W(n_j, s_j, d-1)}{\prod_{i=1}^\ell W(n_i, s_i, d-1)}\\
            &\leq \frac{\growth(n)}{\prod_{i=1}^\ell \growth(n_i)} \cdot \frac{ \sum_{i=1}^\ell s_i \cdot \prod_{j=1}^\ell \growth(n_j)}{\prod_{i=1}^\ell W(n_i, s_i, d-1)}\\
            &= \frac{\growth(n)}{W(n, s, d)} \cdot s
    \end{align*}
    and hence $s \geq L(n, d) \cdot \frac{W(n, s, d)}{\growth(n)}$.
\end{proof}

We now have all the needed tools to prove that monotone formulas of multiplication depth $\dmult$ computing $\imm k n$ require at least $n \cdot k^{\dmult(n^{\sfrac{1}{\dmult}} - 1)}$ leaves.

\begin{proof}[Proof of \cref{thm:depth_snir_imm}]
\hypertarget{proof:depth_snir_imm}{}
    We choose the same weight $w(q) = |\supp(q)|$ and the same growth function $\growth(n) = k^{n-1}$ for $\imm k n$ as \cite{DBLP:conf/icalp/Snir80}.
    We refer to that original paper to show that these fulfill the conditions of a weight and compatible growth function.
    
    To use \cref{thm:snir_general} we need to show that the function
    \[
        L(n, d) = n \cdot k^{d(n^{\sfrac{1}{d}} -1)}
    \]
    fulfills \cref{eq:L_base,eq:L_ind}.
    We explicitly define $L(n, 0) = n$ for $n \leq 1$ and $L(n, 0) = \infty$ for $n \geq 2$.
    Clearly $L(1, d) = 1 = \growth(1)$ for all $d \geq 0$.
    Now let $n = \sum_{i=1}^\ell n_i$ for some $\ell \geq 1$.
    For $d = 1$, if there are any $n_i \geq 2$, then \cref{eq:L_ind} is trivially fulfilled, since the right-hand side will be $\infty$.
    So $n_i = 1$ and $\ell = n$ and we have
    \begin{align*}
        L(n, 1) = n \cdot k^{n-1} = \frac{\growth(n)}{\growth(1)^\ell} \cdot n = \frac{\growth(\sum_{i=1}^\ell n_i)}{\prod_{i=1}^\ell \growth(n_i)} \cdot \sum_{i=1}^\ell L(n_i, 0)\,.
    \end{align*}
    Otherwise, for $d \geq 2$:
    \begin{align*}
        \frac{\growth(\sum_{i=1}^\ell n_i)}{\prod_{i=1}^\ell \growth(n_i)} \cdot \sum_{i=1}^\ell L(n_i, d-1)
            &= \frac{k^{n-1}}{k^{n-\ell}} \cdot \sum_{i=1}^\ell n_i \cdot k^{(n_i^{\sfrac{1}{(d\mathord{-}1)}} - 1)(d-1)}\\
            &= \sum_{i=1}^\ell n_i \cdot k^{n_i^{\sfrac{1}{(d\mathord{-}1)}}(d-1) - d + \ell}\\
            &\geq \ell \cdot (\frac{n}{\ell}) \cdot k^{(\frac{n}{\ell})^{\sfrac{1}{(d\mathord{-}1)}}(d-1) - d + \ell} && \text{(Jensen's inequality)}\\
            &= n \cdot k^{d\left(\frac{\ell + \sum_{i=1}^{d-1}(\frac{n}{\ell})^{\sfrac{1}{(d\mathord{-}1)}}}{d}\right) - d}\\
            &\geq n \cdot k^{d(\ell \cdot \prod_{i=1}^{d-1}(\frac{n}{\ell})^{\sfrac{1}{(d\mathord{-}1)}})^{\sfrac{1}{d}} - d} && \text{(AM-GM inequality)}\\
            &= n \cdot k^{dn^{\sfrac{1}{d}} - d}\\
            &= L(n, d)
    \end{align*}
    We can apply Jensen's inequality because $n_i \cdot k^{n_i^{\sfrac{1}{(d\mathord{-}1)}}(d-1) - d + \ell}$ is convex in $n_i$.
    It becomes an equality for $n_1 = n_2 = \ldots = n_\ell = \frac{n}{\ell}$.
    Furthermore, the AM-GM inequality becomes an equality if $\ell = (\frac{n}{\ell})^{\sfrac{1}{(d\mathord{-}1)}}$, i.e.\ for $\ell = n^{\sfrac{1}{d}}$.
    \cref{thm:snir_general} then finishes the lower bound.
\end{proof}

Another important monotone polynomial is the permanent $\per n = \sum_{\pi \in S_n} \prod_{i=1}^n x_{i, \pi(i)}$.
It is complete for $\ccVNP$ and conjectured not to be in $\ccVP$.
In fact this is one of the alternative formulations of the $\ccVP$ vs $\ccVNP$ conjecture.
In the monotone world, more is known: The permanent provably requires monotone circuits of size at least $n(2^{n-1}-1)$\cite{DBLP:journals/jacm/JerrumS82} and monotone formulas with at least $2^{2n-0.25\log^2(n) + O(\sfrac{1}{n})}$ leaves \cite{DBLP:conf/icalp/Snir80}.

We now give an exact optimal construction for constant multiplication depth monotone formulas for the permanent.
Note that \enquote{optimal construction} in this case only refers to the general shape of the monotone formula, not the exact choices of the recursive parameters.
\begin{theorem}
    \label{thm:depth_snir_per_non_explicit}
    Let $\dmult \geq 2$.
    For every $n$ there is a monotone formula for $\per n$ with multiplication depth $\dmult$ and exactly $L(n,\dmult)$ leaves, and no such formula has fewer leaves, where
    \[
        L(n, d) := \begin{cases}
            n \cdot n! & \text{for $d=1$}\\
            \displaystyle\min_{\substack{n = \sum_{i=1}^\ell n_i}} \binom{n}{n_1,n_2, \ldots, n_\ell} \sum_{i=1}^\ell L(n_i, d-1)& \text{for $d \geq 2$.}
        \end{cases}
    \]
    As usual, the minimum is over all values of $\ell \geq 1$ and all decompositions of $n$ into $n_i \geq 1$.
\end{theorem}
\begin{proof}
    We start with the lower bound.
    We choose the same weight $w(q) = |\supp(q)|$ and the same growth function $\growth(n) = n!$ for $\per n$ as \cite{DBLP:conf/icalp/Snir80}.
    For compatibility, we again refer to the original paper.
    Expanding the domain of $L$ via $L(n, 0) = n$ for $n \leq 1$ and $L(n, 0) = \infty$ for $n \geq 2$, it is easy to see that $L$ satisfies the conditions of \cref{thm:snir_general}.
    Thus, no monotone formula with multiplication depth $\dmult$ computing $\per n$ can have less than $L(n, \dmult)$ leaves.

    It remains to show that this number of leaves is indeed possible by a recursive construction.
    This follows from generalizing the construction of \cite{DBLP:journals/mst/ShamirS80} by using the iterated block-Laplace expansion.
    For $d = 1$, the defining monotone formula $\per n = \sum_{\pi \in S_n} \prod_{i=1}^n x_{i, \pi(i)}$ has multiplication depth $1$ and exactly $L(n, 1) = n \cdot n!$ leaves.
    To construct a monotone formula for $\per n$ with multiplication depth $d \geq 2$, let $\ell$ and $n_1, \ldots, n_\ell$ be such that the minimum in the definition of $L(n, d)$ is attained.
    Let the rows of the symbolic matrix $X$ be partitioned into blocks $R_1, \ldots, R_\ell$ where block $R_i$ has $n_i$ rows.
    The iterated block-Laplace expansion gives us the formula
    \[
        \per n(X) = \sum_{\substack{(C_1, \ldots, C_\ell)\\|C_i| = n_i}}\prod_{i=1}^\ell \per {n_i} (X[R_i, C_i])
    \]
    where the sum is over all ordered partitions of the columns of $X$ and $X[R_i, C_i]$ denotes the submatrix of $X$ with rows $R_i$ and columns $C_i$.
    Using induction to construct monotone formulas of multiplication depth $d-1$ for the $\per {n_i}$, we obtain formulas with $L(n_i, d-1)$ leaves for each of them.
    Moreover, the number of ordered partitions with $|C_i| = n_i$ is given by the multinomial coefficient.
\end{proof}

While \cref{thm:depth_snir_per_non_explicit} gives an exact lower bound, it doesn't give a particularly simple characterization of the optimal number of leaves, which we address now.

\begin{theorem}
    \label{thm:depth_snir_per}
    Let $\dmult \geq 2$ be fixed.
    The optimal number of leaves in any monotone formula of multiplication depth $\dmult$ computing the permanent $\per n$ is $\left(\left(e+o(1)\right) \ln^{\circ(\dmult-1)}(n)\right)^n$.
\end{theorem}
\begin{proof}
    Let $L$ be the function from \cref{thm:depth_snir_per_non_explicit}.
    We show $L(n, \dmult)^{\sfrac{1}{n}} = \left(e+o(1)\right) \ln^{\circ(\dmult-1)}(n)$ via induction on $d$.
    The proof consists of two parts, first we will prove that almost balanced partitions already yield the optimal behavior, allowing us to turn the minimization into just a single variable optimization.
    Secondly, we will solve the resulting single variable optimization problem using our induction.
    Throughout, we will use that $L$ is monotone in $n$.
    
    Let $\dmult \geq 2$ and let $n = \sum_{i=1}^\ell n_i$ be a partition that minimizes $\frac{n!}{\prod_{i=1}^\ell n_i!} \sum_{i=1}^\ell L(n_i, \dmult-1)$ and let $m = \max_{1 \leq i \leq \ell}n_i$ be its maximum element.
    Then%
    \begin{align*}
        L(n, \dmult)^{\sfrac{1}{n}} &= \left(\frac{n!}{\prod_{i=1}^\ell n_i!} \sum_{i=1}^\ell L(n_i, \dmult-1)\right)^{\sfrac{1}{n}}\\
        &\geq \left(\frac{n!}{\prod_{i=1}^\ell n_i!} L(m, \dmult-1)\right)^{\sfrac{1}{n}}\\
        &\geq \left(\frac{n!}{(m!)^{\sfrac{n}{m}}} L(m, \dmult-1)\right)^{\sfrac{1}{n}} && \text{(Monotonicity of $(k!)^{\sfrac{1}{k}}$)}\\
        &= \frac{(n!)^{\sfrac{1}{n}}}{(m!)^{\sfrac{1}{m}}} L(m, \dmult-1)^{\sfrac{1}{n}}
    \end{align*}
    On the flip side, for any $m \geq 1$, splitting $n$ into $\ell = \lceil\frac{n}{m}\rceil$ almost equal pieces gives us a partition where all components are either $m$ or $m+1$.
    
    \begin{align*}
        L(n, \dmult)^{\sfrac{1}{n}} & \leq \left(\frac{n!}{\prod_{i=1}^\ell n_i!} \sum_{i=1}^\ell L(n_i, \dmult-1)\right)^{\sfrac{1}{n}}\\
        &\leq \left(\frac{n!}{(m!)^\ell} \cdot \ell \cdot L(m+1, \dmult-1)\right)^{\sfrac{1}{n}}\\
        &\leq \left(\frac{n!}{(m!)^\ell} \cdot \ell \cdot L(m+1, \dmult-1)\right)^{\sfrac{1}{n}}\\
        &= \left(\frac{n!}{(m!)^\ell}\right)^{\sfrac{1}{n}} \cdot \ell^{\sfrac{1}{n}} \cdot L(m+1, \dmult-1)^{\sfrac{1}{n}}\\
        &= \left(\frac{n!}{(m!)^\ell}\right)^{\sfrac{1}{n}} \cdot \ell^{\sfrac{1}{n}} \cdot \frac{L(m+1, \dmult-1)^{\sfrac{1}{n}}}{L(m, \dmult-1)^{\sfrac{1}{n}}} \cdot L(m, \dmult-1)^{\sfrac{1}{n}}\\
        &= \left(\frac{n!}{(m!)^\ell}\right)^{\sfrac{1}{n}} \cdot \ell^{\sfrac{1}{n}} \cdot (1+o(1)) \cdot L(m, \dmult-1)^{\sfrac{1}{n}} && \text{(Induction hypothesis)}\\
        &= \left(\frac{n!}{(m!)^\ell}\right)^{\sfrac{1}{n}} \cdot (1+o(1)) \cdot L(m, \dmult-1)^{\sfrac{1}{n}} && \text{($\lim_{n\to\infty}\left\lceil\frac{n}{m}\right\rceil^{\sfrac{1}{n}} = 1$)}\\
        &\leq (1+o(1)) \frac{(n!)^{\sfrac{1}{n}}}{(m!)^{\sfrac{1}{m}}} L(m, \dmult-1)^{\sfrac{1}{n}}
    \end{align*}
    Note that for $\dmult = 2$, we cannot apply the induction hypothesis itself, but still have 
    \begin{align*}
        \lim_{n\to \infty}\frac{L(m+1, \dmult-1)^{\sfrac{1}{n}}}{L(m, \dmult-1)^{\sfrac{1}{n}}} = \left(\frac{(m+1)(m+1)}{m}\right)^{\sfrac{1}{n}} = 1\,.
    \end{align*}
    Combining both inequalities with Stirling's approximation, we get
    \begin{align*}
        L(n, \dmult)^{\sfrac{1}{n}} = (1+o(1)) \min_{m \geq 1} \frac{(n!)^{\sfrac{1}{n}}}{(m!)^{\sfrac{1}{m}}} L(m, \dmult-1)^{\sfrac{1}{n}} = (1+o(1)) \min_{m \geq 1} \frac{n}{m} \left(L(m, \dmult-1)^{\sfrac{1}{m}}\right)^{\sfrac{m}{n}}\,.
    \end{align*}
    If $\dmult = 2$, then $L(m, \dmult-1)^{\sfrac{1}{m}} = (1+o(1))\frac{m}{e}$ by Stirling's approximation.
    Setting $x = \frac{n}{m}$,
    \begin{align*}
        L(n, \dmult)^{\sfrac{1}{n}} = (1+o(1)) \min_{x \geq 1} x e^{(o(1) + \ln(n) - \ln(x) - 1) \frac{1}{x}}\,.
    \end{align*}
    Taking derivatives, this expression is minimized for $x = (1+o(1))\ln(n)$, giving $L(n, \dmult)^{\sfrac{1}{n}} = (e+o(1))\ln(n)$.
    Otherwise, for $\dmult > 2$, using the induction hypothesis and doing the same substitution $x = \frac{n}{m}$,
    \begin{align*}
        L(n, \dmult)^{\sfrac{1}{n}} &= (1+o(1)) \min_{x \geq 1} x \left(\left(e+o(1)\right) \ln^{\circ(\dmult-2)}\left(\frac{n}{x}\right)\right)^{\sfrac{1}{x}}\\
        &= (1+o(1)) \min_{x \geq 1} x e^{\ln\left(\left(e+o(1)\right) \ln^{\circ(\dmult-2)}\left(\frac{n}{x}\right)\right)\frac{1}{x}}\\
        &= (1+o(1)) \min_{x \geq 1} x e^{\left(1 + o(1) + \ln^{\circ(\dmult-1)}\left(\frac{n}{x}\right)\right)\frac{1}{x}}\,.
    \end{align*}
    Similarly, this expression is minimized for $x = (1+o(1)) \ln^{\circ(\dmult-1)}(n)$, resulting in $L(n, \dmult)^{\sfrac{1}{n}} = \left(e+o(1)\right) \ln^{\circ(\dmult-1)}(n)$.
\end{proof}

\subsection{Indirect Lower Bounds}
While \cref{thm:snir_general} only gives lower bounds on monotone constant-depth algebraic formulas, we indirectly also get lower bounds for different monotone models of computation, in particular for monotone constant-depth algebraic circuits.

To obtain lower bounds for constant-depth algebraic circuits, we use the well known unfolding construction turning a size\footnote{In this context, we consider the size of a circuit to be the number of edges.} $s$ and depth $d$ monotone algebraic circuit into a monotone algebraic formula with at most $s^d$ leaves and depth $d$.
Additionally, this construction preserves the multiplication depth.

Combined with \cref{thm:snir_general} we obtain the general:
\begin{corollary}
    Let $p$ be a monotone polynomial of degree $n$ and let $L$ be as in \cref{thm:snir_general}.
    Then, any arbitrary-fan-in monotone circuit of depth $d$ and multiplication depth $\dmult$ that computes $p$ requires size at least $(L(n, \dmult))^{\sfrac{1}{d}}$.
\end{corollary}

In particular, \cref{thm:depth_snir_imm} gives us the following.
\begin{corollary}
    \label{cor:circuit_depth_snir_imm}
    Let $C$ be an arbitrary-fan-in monotone circuit of depth $d$ and multiplication depth $\dmult$ for some $\imm k n$.
    Then $C$ has size at least $n \cdot k^{\frac{\dmult}{d}(n^{\sfrac{1}{\dmult}} - 1)}$.
\end{corollary}
This improves on the previously known lower bound of $k^{\Omega\left(\frac{1}{d} n^{\sfrac{1}{\lceil \sfrac{d}{2} \rceil}}\right)}$ by \cite{JAJA19851155}.
Similarly, \cref{thm:depth_snir_per} implies:
\begin{corollary}
    \label{cor:circuit_depth_snir_per}
    Let $\dmult \geq 2$ be fixed.
    The optimal number of leaves in any monotone circuit of depth $d$ and multiplication depth $\dmult$ computing the permanent $\per n$ is 
    $$\left(\left(e+o(1)\right) \ln^{\circ(\dmult-1)}(n)\right)^{\sfrac{n}{d}}.$$
\end{corollary}

\section{Computing Iterated Matrix Multiplications by Narrow mABPs}
We have already seen in \cref{lem:imm_to_depth_formula} how to compute $\imm k n$ by a monotone formula of addition depth $\dadd$ of size $O(n \cdot k^{\dadd(\lceil n^{\sfrac{1}{\dadd}} \rceil-1)})$ and in \cref{cor:formulatoabp} that this in turn can then be computed by an mABP of width $2\dadd + 1$ and depth $O(n \cdot k^{\dadd(\lceil n^{\sfrac{1}{\dadd}} \rceil-1)})$.
Setting $\dadd = \lfloor\frac{w-1}{2}\rfloor$ we get
\begin{corollary}
    Let $k \geq 2$ and $w \geq 3$.
    Then $\imm k n$ can be computed by an mABP of width $w$ and depth $O(n \cdot k^{\lfloor\frac{w-1}{2}\rfloor(\lceil n^{\sfrac{1}{\lfloor\frac{w-1}{2}\rfloor}} \rceil-1)})$.
\end{corollary}

Considering the gap to the lower bound of $\left(\frac{k}{w-1}\right)^{\frac{1}{2}\left\lfloor n^{\frac{1}{w-2}} \right\rfloor - 1}$ on the depth, given by \cref{thm:main_separation}, this begs the question whether this upper bound can be improved.
By \cref{thm:depth_snir_imm}, the intermediate constant-depth formula has optimal size.
Any conversion of arbitrary addition depth $\dadd$ formula of size $s$ to a width $w$ mABP requires depth $\Omega(\frac{s}{w^2})$ by simple counting arguments, so any better construction needs to reuse intermediate results more.

In this section we answer this question in the positive, by using the fact that mABPs are not just efficient at multiplying individual numbers, but they are also efficient at multiplying matrices of bounded (by the width of the mABP) size.

We say a multi-input, multi-output mABP $A$ has $\sigma$ sources and $\tau$ sinks if the first layer contains the nodes $s_1, \ldots, s_\sigma$ and the last layer contains the nodes $t_1, \ldots, t_\tau$.
It computes a $\sigma \times \tau$ matrix $F$ of polynomials with $F^{(A)}_{i,j} = \fpartial{s_i}{t_j}{A}$.
We can similarly to \cref{lem:abpprod,lem:abpsum} add and multiply the $F^{(A)}$ (as matrices) using mABPs.
As previously, we incur a width penalty for addition.
\begin{lemma}
    \label{lem:multiabpprod}
    Let $A_1, \ldots, A_r$ be multi-input, multi-output mABPs of depths $d_1, \ldots, d_r$ and widths $w_1, \ldots, w_r$ with $\sigma_1, \ldots, \sigma_r$ sources and $\tau_1, \ldots, \tau_r$ sinks.
    If $\sigma_{i+1} = \tau_i$ for all $i \in \{1, \ldots, r-1\}$, then there is an mABP $A$ of depth $\sum_{i=1}^r d_i$ and width $\max\{w_1, \ldots, w_r\}$, computing $\prod_{i=1}^r F^{(A_i)}$.
\end{lemma}
\begin{proof}
    Identifying the sources of $A_{i+1}$ with the sinks of $A_i$ for all $i \in \{1, \ldots, r-1\}$ immediately gives us the desired mABP $A$.
\end{proof}

\begin{lemma}
    \label{lem:multiabpsum}
    Let $A_1, \ldots, A_r$ be multi-input, multi-output mABPs of depths $d_1, \ldots, d_r$ and widths $w_1, \ldots, w_r$, all with $\sigma$ sources and $\tau$ sinks.
    Then there is an mABP $A$ of depth $\sum_{i=1}^r d_i$ and width $\sigma + \tau + \max\{w_1, \ldots, w_r\}$, computing $\sum_{i=1}^r F^{(A_i)}$.
\end{lemma}
\begin{proof}
    We follow mostly along \cref{lem:abpsum}.
    We construct $A$ from $\sigma$ paths $s^{\kappa}_0, \ldots, s^{\kappa}_{\ell_1}$ from layer $0$ to layer $\ell_1 = \sum_{i=1}^{r-1}d_i$ and $\tau$ paths $t^\lambda_{d_1}, \ldots, t^\lambda_{\ell_2}$ from layer $d_1$ to layer $\ell_2 = \sum_{i=1}^r d_i$.
    All edges along these $\sigma+\tau$ many paths have weight $1$.
    For $1 \leq j \leq r$ we now place $A_j$ in such a way in $A$ that its $\kappa$-th source is identified with $s_{\sum_{i=1}^{j-1}d_i}$ and its $\lambda$-th sink is identified with $t_{\sum_{i=1}^j d_i}$.

    We observe that $A$ is indeed a correctly formed mABP as each of the $A_j$ has depth $d_j$ and its sources and sinks are identified with some $s_{\sum_{i=1}^{j-1}d_i}$ and $t_{\sum_{i=1}^j d_i}$, which are in layers that are $d_j$ apart.
    The width of $A$ is at most $\sigma + \tau + \max\{w_1, \ldots, w_r\}$ due to the $A_j$ being in disjoint layers except their sources and sinks.
    $A$ clearly computes $\sum_{i=1}^r F^{(A_i)}$: Any path starting from $s^{\kappa}_0$ and ending in $t^{\lambda}_{\ell_2}$ has to cross exactly one of the $A_j$ from its $\kappa$-th source to its $\lambda$-th sink.
\end{proof}

Using a construction similar to \cref{lem:imm_to_depth_formula}, we get a slightly more efficient construction to represent $\imm k n$ with an mABP of width $w < k$.

\begin{theorem}
\label{thm:narrow-ABPs-width-to-depth}
    Let $w \geq 3$.
    Then, for every $r \in \{1, \ldots, \lfloor\frac{w-1}{2}\rfloor\}$, the polynomial $\imm k n$ can be computed by an mABP of width $w$ and depth at most $n \cdot \left\lceil \sfrac{k}{\left\lfloor\frac{w-2}{(2r-1)}\right\rfloor}\right\rceil^{r(\lceil n^{\sfrac{1}{r}} \rceil - 1)}$.
\end{theorem}
\begin{proof}
    Let $k$ be fixed.
    Our plan is to construct multi-input and multi-output mABPs, computing the family of polynomial matrices
    \[
        F_{I,J, N, \rho,i} = (\bX^{(i)} \cdot \bX^{(i+1)} \cdots \bX^{(i+N^\rho-1)})[I,J],
    \]
    where $I, J \subseteq \{1, \ldots, k\}$ and, as usual, $\bX^{(j)}$ are $k \times k$ matrices of variables.
    In the end, $\imm k {N^r}$ can be directly retrieved from $F_{\{1\}, \{1\}, N, r, 1}$ while for general $\imm k n$ we construct $\imm k {N^r}$ for $N = \lceil n^{\sfrac{1}{r}} \rceil$ and project away the surplus matrices.

    For $\rho = 0$, we have $F_{I,J, N, \rho,i} = \bX^{(i)}[I, J]$ which can be computed by an mABP with $|I|$ sources and $|J|$ sinks and no intermediate layers, i.e.\ depth $1$.
    For $\rho \geq 1$, let $b = \left\lceil \sfrac{k}{\left\lfloor\frac{w-2}{(2r-1)}\right\rfloor}\right\rceil$ and let $P_1, \ldots, P_b$ be a partition of $\{1, \ldots, k\}$ where no part is bigger than $\frac{w-2}{2r-1}$ elements.
    We decompose
    \begin{align}
        F_{I,J,N,\rho,i} = \sum_{K_0, \ldots, K_N}\prod_{j=1}^{N}F_{K_{j-1}, K_j, N, \rho-1,i+(j-1) \cdot N^{\rho-1}} \label{eq:imm_to_smaller_imm_formula:recursion}
    \end{align}
    where the sum is over all $K_0, \ldots, K_N$, s.t.\ $K_0 = I$, $K_N = J$ and all other $K_j$ are one of $P_1, \ldots, P_b$.

    A simple induction along this decomposition, using \cref{lem:multiabpsum,lem:multiabpprod}, shows that for $\rho \geq 1$, $F_{I,J,N,\rho,i}$ has multi-input and multi-output mABPs of width $\leq |I|+|J|+(2\rho - 1)\cdot \frac{w-2}{2r-1}$ and depth $\leq N^{\rho} \cdot b^{\rho(N-1)}$.

    In the end, for $I = J = \{1\}$, $\rho = r$ and $N=\lceil n^{\sfrac{1}{r}} \rceil$, we get a single-source and single-output mABP of width $w$ and depth $\lceil n^{\sfrac{1}{r}} \rceil^\rho \cdot \left\lceil \sfrac{k}{\left\lfloor\frac{w-2}{(2r-1)}\right\rfloor}\right\rceil^{r(\lceil n^{\sfrac{1}{r}} \rceil - 1)}$ computing $\imm {k} {N^r}$ and thus $\imm k n$.
    
    Again, as in \cref{lem:imm_to_depth_formula}, we can reduce the factor of $\lceil n^{\sfrac{1}{r}} \rceil^\rho$ to just $n$, by choosing the recursive sub-polynomials in \cref{eq:imm_to_smaller_imm_formula:recursion} non-uniformly, thus avoiding surplus matrices directly.
\end{proof}

\section{Acknowledgements}

We thank Prerona Chatterjee and Utsab Ghosal for the fruitful discussion during the WACT 2026 and for making us aware of some relevant literature.

\section{AI Methodology}

No AI was used to find any of the ideas that led to our results, except for an initial idea for \cref{lem:logvslin} by ChatGPT 5.5, which the authors then expanded to \cref{cor:coarse_abp_separation_logf_vs_logg}.

ChatGPT 5.5 was used to assist with optimizing the specific parameters of the recursions in \cref{thm:main_separation}.
Moreover, ChatGPT 5.6 Sol was used to assist with finding simple closed forms for the function $L$, subject to the constraints of \cref{thm:snir_general}, in \cref{thm:depth_snir_imm,thm:depth_snir_per}.
Finally, Qwen 3.8 Flash Next identified a couple of minor edge cases which were subsequently fixed by the authors.

\appendix

\section{Simplification}
\label{sec:simplify}
Many of our proofs hinged on being able to count the monomials in an mABP.
Ideally, every $s$-$t$-path would contribute at most one monomial to the resulting polynomial. 
Of course, this is not true, even in the monotone setting. 
We therefore introduce a restricted class of mABPs that will allow for more straightforward counting.
We remind of the individual criteria that make an mABP simple:
\begin{definition}[Simple ABPs]
    \label{def:simple}
    An mABP $A$ is simple if all of the following are satisfied:
    \begin{enumerate}
        \item There is no vertex $v$ with $\fpartial{s}{v}{A} = 0$ or $\fpartial{v}{t}{A} = 0$.
        \item In every layer, there are no two vertices $u, v$ with $\fpartial{s}{u}{A} = c \cdot \fpartial{s}{v}{A}$ or $\fpartial{u}{t}{A} = c \fpartial{v}{t}{A}$ for some constant $c \neq 0$.
        \item Every edge in $A$ is labeled by a constant or a multiple of a single variable.
    \end{enumerate}
\end{definition}
For a simple mABP $A$, it is easy to see that we can indeed bound $|\supp(\ffull{A})|$ by the number of $s$-$t$-paths in $A$. In the remainder of the section, we will proceed to show that simplicity is not a real restriction.

\begin{lemma}
    \label{lem:linear_deps}
    Let $A$ be a width $w$, depth $d$ mABP. Then there is an mABP $A'$ with width $w$ and depth at most $d$ that satisfies properties $1.$ and $2.$ in \cref{def:simple} with $\ffull{A} = \ffull{A'}$.
\end{lemma}
\begin{proof}
    We first remove all vertices $v$ with $\fpartial{s}{v}{A} = 0$ or $\fpartial{v}{t}{A} = 0$, which does not change the polynomial.

    Now, suppose that there is a layer with vertices $v_1, v_2$ such that $\fpartial{s}{v_1}{A} = c \fpartial{s}{v_2}{A}$ with $c\neq 0$. Since $A$ is monotone, both polynomials are monotone, so we can assume that $c > 0$.

    Consider some vertex $u_j$ in the next layer and let $\ell_{i,j}$ be the label on the edge from $v_i$ to $u_j$ (we consider non-existing edges to have $\ell_{i,j} = 0$).
    We completely remove the vertex $v_1$ and replace all labels $\ell_{2,j}$ by $\ell_{2,j} + c \ell_{1,j}$. 
    
    This way, we obtain the mABP $A'$. The polynomial is unchanged since
    \begin{align*}
        \ffull{A} 
        &= \sum_j (\fpartial{s}{v_1}{A} \cdot \ell_{1,j} \cdot \fpartial{u_j}{t}{A}
        + \fpartial{s}{v_2}{A} \cdot \ell_{2,j} \cdot \fpartial{u_j}{t}{A})
        + g
        \\&= \sum_j ( c \fpartial{s}{v_2}{A} \cdot \ell_{1,j} \cdot \fpartial{u_j}{t}{A} +  \fpartial{s}{v_2}{A} \cdot \ell_{2,j} \cdot \fpartial{u_j}{t}{A})
        + g
        \\&= \sum_j (\fpartial{s}{v_2}{A} \cdot (\ell_{2,j} + c \ell_{1,j}) \cdot \fpartial{u_j}{t}{A}) + g = \ffull{A'}
    \end{align*}
    for some polynomial $g$.

    We handle linear dependencies between $\fpartial{v_i}{t}{A}$ analogously, but replacing the incoming edges instead of outgoing edges. The modifications do not introduce any subtractions, so the resulting mABP remains monotone.
\end{proof}

The next lemma is a crucial observation for mABPs: If we assume that the final polynomial is homogeneous, then any \enquote{intermediate} polynomials must have been homogeneous as well, as it is not possible to cancel any intermediate inhomogeneous monomials in the monotone setting.

\begin{lemma}
    \label{lem:intermediate_homogeneous}
    If an mABP $A$ satisfying property $1.$ in \cref{def:simple} computes a homogeneous polynomial, then 
    for every vertex $v$, both $\fpartial{s}{v}{A}$ and $\fpartial{v}{t}{A}$ are homogeneous and all edges are labeled by homogeneous polynomials of degree at most $1$.
\end{lemma}
\begin{proof}
    Let $v$ be a vertex such that $\fpartial{s}{v}{A}$ or $\fpartial{v}{t}{A}$ are not homogeneous. 
    Then $\fpartial{s}{v}{A} \cdot \fpartial{v}{t}{A}$ is also not homogeneous since neither of them can be $0$ (property $1.$). 
    Since $A$ is monotone, every monomial appearing in $\fpartial{s}{v}{A} \cdot \fpartial{v}{t}{A}$ will also have to appear in $\ffull{A}$, a contradiction to $\ffull{A}$ being homogeneous.
    Now let $\ell$ be a label of an edge connecting two nodes $u$ and $v$. 
    By the same argument $\fpartial{s}{u}{A} \cdot \ell \cdot \fpartial{v}{t}{A}$ has to be homogeneous, so $\ell$ has to be homogeneous itself.
\end{proof}

This means that any edge is labeled either by a constant, or by a linear combination of single variables. While this is still weaker than the third criterion of simplicity, if we go further and assume that $A$ computes a polynomial with a subsupport of $\imm k n$, we can prove the stricter statement.
\begin{lemma}
    \label{lem:edges_const_or_single_var}
    If an mABP $A$ satisfying property $1.$ in \cref{def:simple} computes a polynomial $\ffull{A}$ with $\supp(\ffull{A}) \subseteq \supp(\imm{k}{n})$, then every edge is labeled either by a constant or a multiple of a single variable.
\end{lemma}
\begin{proof}
    Let $\ell$ be the label on an edge from $u$ to $v$. Since $\ffull{A}$ is homogeneous, we can assume by \cref{lem:intermediate_homogeneous} that $\ell$ is homogeneous of degree $0$ or $1$. Without loss of generality, assume that $\ell$ is a linear combination of exactly two distinct variables $x_{\kappa, \lambda}^{(i)}$ and $x_{\kappa', \lambda'}^{(i')}$ (the case of larger linear combinations is fully analogous), so $\ell = c_1 x_{\kappa, \lambda}^{(i)} + c_2 x_{\kappa', \lambda'}^{(i')}$.

    Consider some $s$-$t$-path $p$ through $\ell$. Due to property $1.$, we know that it exists and its weight is $\ell \cdot g \neq 0$ for some polynomial $g$.

    We will first show that $i = i'$. Since $A$ is monotone and $\ffull{A}$ is set-multilinear (with respect to the matrices $\bX^{(1)}, \ldots, \bX^{(n)}$), so is $\ell \cdot g = c_1 x_{\kappa, \lambda}^{(i)} \cdot g + c_2 x_{\kappa', \lambda'}^{(i')} \cdot g$. Its monomials containing exactly one variable from every matrix is only possible if $i = i'$.

    Now we show that $\kappa = \kappa'$. If $i = 1$, this is evident as only $\kappa = \kappa' = 1$ appears in the support.
    Now, assume that $i > 1$ and $\kappa \neq \kappa'$. Then, $\ell \cdot g$ must contain a variable $x_{\mu, \kappa}^{(i-1)}$ from the previous matrix. But then $\ell \cdot g$ also contains a monomial $m$ with both $x_{\mu, \kappa}^{(i-1)}$ and $x_{\kappa', \lambda'}^{(i)}$. But $m \in \supp(\ell \cdot g) \subseteq \supp(\ffull{A}) \subseteq \supp(\imm{k}{n})$ due to monotonicity, which is a contradiction since $\kappa \neq \kappa'$ and $\imm{k}{n}$ has no monomial with $x_{\mu, \kappa}^{(i-1)} x_{\kappa', \lambda'}^{(i)}$ as a factor. A symmetric argument shows that $\lambda = \lambda'$.
\end{proof}

Combining the above, \cref{lem:simplicity} directly follows from \cref{lem:linear_deps} and \cref{lem:edges_const_or_single_var}.

\nolinenumbers
\printbibliography

\end{document}